\documentclass[11pt,a4paper]{article}
\usepackage{jcappub}
\usepackage{multirow}
\usepackage{graphicx}
\usepackage{amsmath,amsfonts,amsthm}
\usepackage{bm}
\usepackage[normalem]{ulem}

\definecolor{Jose}{RGB}{102,0,0}

\newcommand{\ie}{\textit{i.e.}}
\newcommand{\eg}{\textit{e.g.}}

\newcommand{\change}[1]{#1}

\begin{document}

\title{Towards a more rigorous treatment of uncertainties on the velocity distribution of dark matter particles for capture in stars}

\author[a]{Jos\'{e} Lopes,}
 \affiliation[a]{Centro de Astrof\'isica e Gravita\c{c}\~{a}o—CENTRA, Departamento de F\'isica, Instituto Superior T\'{e}cnico—IST,\\
 Universidade de Lisboa—UL, Av. Rovisco Pais 1, 1049-001 Lisboa, Portugal}
\author[b]{Thomas Lacroix}%
\affiliation[b]{Instituto de F\'isica Te\'orica UAM/CSIC, Universidad Aut\'onoma de Madrid, 28049 Madrid, Spain}%
\author[a]{and Il\'{i}dio Lopes}

\emailAdd{josevlopes@tecnico.ulisboa.pt}
\emailAdd{thomas.lacroix@uam.es}
\emailAdd{ilidio.lopes@tecnico.ulisboa.pt}

\date{\today}

\abstract{Dark matter (DM) capture in stars offers a rich phenomenology that makes it possible to probe a wide variety of particle DM scenarios in diverse astrophysical environments. In spite of decades of improvements to refine predictions of capture-related observables and better quantify astrophysical and particle-physics uncertainties, the actual impact of the Galactic phase-space distribution function of DM has been overlooked. In this work, we tackle this problem by making use of self-consistent equilibrium phase-space models based on the Eddington inversion formalism and an extension of this method to a DM halo with some degree of anisotropy in velocity space. We demonstrate that incorrectly accounting for the variation of the DM velocity distribution with position in the Galaxy leads to a systematic error between a factor two and two orders of magnitude, depending in particular on the target star, the DM candidate mass and the type of interaction involved. Moreover, we show that underlying phase-space properties, such as the anisotropy of the velocity tensor, actually play an important part---previously disregarded---and can have a sizable impact on predictions of capture rates and subsequent observables. We argue that Eddington-like methods, which self-consistently account for kinematic constraints on the components of the Galaxy, actually provide a reliable next-to-minimal approach to narrow down uncertainties from phase-space modeling on predictions of observables related to DM capture in stars.}

\maketitle


\section{Introduction}
\label{sec:intro}

In the presence of non-gravitational interactions between dark matter (DM) and the standard sector, DM particles which permeate the Galactic halo can lose energy via elastic scattering with stellar matter, and wind up trapped in the gravitational potential wells of stars. This process is called DM capture, and has been extensively studied in the literature (\eg~\cite{PressSpergel1985,Gould1987,lopes2011,catena2016,busoni2017,Dasgupta2019, Garani2019}) due to its ubiquity in studies that concern the effects, manifestations and/or signals of DM in stars. If they lose enough energy through elastic scattering with stellar matter and settle down in the star's interior, DM particles may affect the well-known physical processes that govern stellar physics. This gives rise to a rich new phenomenology which can be used to explore the nature of DM itself. 

In the Sun, interactions of weakly interacting massive particles (WIMPs) captured in the solar core were first proposed to address the solar neutrino problem \cite{steigman1978,spergel85,faulkner1985}. Today we know that the deficit observed in the solar neutrino flux is instead caused by neutrino oscillations. In spite of this, the effects of DM energy transport in stars have been extensively studied in other contexts \cite{renzini1987,bouquet1989,dearborn1989,gould1990}, and have been used to both constrain DM--nucleon interactions and study their implications for the standard picture of astrophysics and stellar evolution \cite{lopes2010,Taoso2010,Iocco2012,casanellas2013,scott2015,lopes2019}. In these various contexts, the main ingredient is the capture rate which determines the magnitude of the effects DM can have on a given star, and which needs to be modeled as accurately as possible.

An additional consequence of DM trapping in stars, for which an accurate treatment of the capture rate is equally relevant, is self-annihilation of captured DM particles and the resulting observational consequences of its by-products. For example, the annihilation of WIMPs captured in the Sun creates a steady flux of neutrinos that would escape the solar plasma essentially unscathed \cite{Silk1985,gaisser1986,srednicki1987}. The absence of such a signal in neutrino telescopes sets stringent upper bounds on the DM annihilation rate in the Sun \cite{choi2015,aartsen2017} which can only be translated into a limit on the DM-nucleon scattering cross section through accurate predictions for the capture rate. 

Beside neutrinos, which promptly escape the star, DM annihilates to other SM particles which ultimately lose energy within the stellar plasma, behaving as a \textit{de-facto} source of energy which can compete with---and in some cases overtake---standard nuclear reactions \citep{salati1989,moskalenko2007}. The additional energy injection caused by the annihilation of captured DM particles can have an effect on the appeareance of entire globular clusters \cite{casanellas2011}, extend the lifetime of early-type stars, or even give rise to an entirely new stellar evolution stage---referred to as dark stars in the literature---in which stability is achieved through the balance of DM annihilation and self-gravity \cite{spolyar2008,iocco2008,yoon2008,taoso2008,freese2008,fairbairn2008}.

Capture of DM particles can also have important consequences for compact objects such as white dwarfs (WDs) and neutron stars (NSs). On the one hand, energy production from (WIMP-like) DM annihilation can heat the degenerate core of these compact objects and alter their cooling process \cite{Kouvaris2008,Bertone2008,Kouvaris2010,deLavallaz2010}. On the other hand, DM annihilation can be negligible in classes of theoretically motivated models such as asymmetric DM (\eg~\cite{kaplan2009}). This would for instance cause particles to accumulate freely inside a NS. Then, if the total mass of DM particles captured by a NS embedded in a DM halo is sufficiently large for the resulting self-gravity to overcome the neutron degeneracy pressure, the NS should collapse, thus turning an otherwise stable star into a black hole (\eg~\cite{goldman1989,gould90,deLavallaz2010,McDermott2012,Guver2014,Garani2019}).

The potential of the capture phenomenology described above as a DM probe can only be fulfilled if predictions for the DM capture rate are sufficiently accurate and all the associated uncertainties are properly accounted for. DM capture occurs if a DM particle from the halo interacts with a nucleon in the stellar plasma and the speed of the DM particle after scattering is lower than the local escape speed in the star. The probability for this event to occur is determined, among other factors, by the frequency with which DM particles traverse the star, and the initial velocity of the DM particle, encoded in the phase-space distribution function (PSDF) $f(\vec{r},\vec{v})$ of the DM population in a galactic halo. Simply put, larger DM densities will increase the probability for particles to traverse the star. Moreover, DM particles with low velocities relative to the target nuclei will be more easily captured. It should be noted that the DM density does not only determine the capture rate through a global normalization, but it is also strongly correlated with the PSDF, a fact that has been essentially overlooked in the literature. As a result, the DM PSDF will have a non-trivial effect on the total capture rate and must therefore be carefully modeled as it is an important source of astrophysical uncertainties.

Several approaches have been developed to estimate the DM PSDF in the context of DM searches, mostly for direct detection in underground laboratories. The most common one used in the literature to predict capture rates in the Sun, the Earth or NSs, relies on the assumption that DM particles in the Galactic halo follow a simple Maxwell-Boltzmann (MB) velocity distribution function (DF) with a velocity dispersion specified in an ad hoc way, usually from the virial theorem (\eg~\cite{Silk1985,PressSpergel1985,Gould1987,Gould1992,Wikstroem2009,Peter2009,Ellis2010,SK2011,Rott2011,IceCube2013,Bernal2013,Rott2013,Boliev2013,ANTARES2013,Busoni2013,Baratella2014,Fornengo2017,Kouvaris2008,Kouvaris2010,deLavallaz2010,McDermott2012,Guver2014,Bramante2013,Bertoni2013,Bramante2014,Bramante2017,Bell2019,Garani2019,Bertone2008,Bell2013,BramanteLinden2014,Bramante2018,Gresham2019,Dasgupta2019}). This has led to the adoption of the so-called standard halo model (SHM), in which the most probable speed is fixed at the value corresponding to the Solar neighborhood, $\sim 220\, \rm km\, s^{-1}$, everywhere in the Galaxy or sometimes adjusted depending on the region of interest, but again in a rather ad hoc way. Although this approach is suitable for simple estimates especially at the Sun's location, the underlying assumption of an isothermal halo leads to a velocity distribution that provides a poor global picture of gravitational systems, as known for a long time from theoretical arguments \cite{King1966,BinneyEtAl2008}, and confirmed using numerical simulations \cite{Kazantzidis2004,Sanchis2004,Wojtak2008}.

Another approach used to estimate the DM PSDF relies on extrapolating outputs of zoom-in cosmological simulations of Milky-Way analogues \cite{Helmi2002,Lavalle2008,Vogelsberger2009,Ling2010,Mao2013,Pillepich2014,Bozorgnia2016,Bozorgnia2019}, with specific applications to DM capture in refs.~\cite{Choi2014,Nunez2019}. However, although numerical simulations are crucial to shed light on the formation, evolution, and structure of galaxies, they may not be the most suitable tool to assess astrophysical uncertainties in a specific object like the Milky Way (MW) with particular characteristics that can be constrained from observations. This critically depends on how typical the MW is with respect to simulated MW-like galaxies.

Ultimately, the main problem is that the previous methods do not account for kinematic constraints on the galactic structure of interest. This is becoming crucial in the era of the \textit{Gaia} space mission \cite{Gaia2016,Gaia2018}, which has been tightening kinematic constraints on the MW, thus leading to more accurate mass models \cite{Cautun2020}. In this context, a data-driven approach has been developed which by construction does account for kinematic constraints since in that case the DM velocity DF is inferred from suitable stellar tracers \cite{Herzog-Arbeitman2018a,Herzog-Arbeitman2018b,Necib2019a}. This method makes it possible to account for non-equilibrium features in the DM velocity DF \cite{Evans2018,O'Hare2018,Necib2019b,Bozorgnia2019,O'Hare2020}, which could have a significant impact on DM observables in the Solar neighborhood. However, this approach is as yet mostly restricted to a few kpc from the Sun, and does not provide a full picture of the DM phase-space in the Galaxy. Yet, full phase-space models are precisely what is missing in the context of capture-related DM searches in the MW.

Although some alternative approaches have been developed to somewhat marginalize over astrophysical uncertainties for local DM searches without a priori having to account for kinematic constraints \cite{Blennow2015,Ibarra2018}, they hardly provide a global dynamical picture of the Galaxy. Instead, other strategies have been developed to predict the DM PSDF throughout galactic structures from first principles, consistently with kinematic measurements. These methods involve solving the steady-state collisionless Boltzmann equation, and have been used essentially in the context of direct and indirect searches. They differ in terms of assumptions and complexity, and range from the Eddington inversion approach, in which one can predict the DM PSDF from the knowledge of the underlying DM density and total gravitational potential \cite{Eddington1916,BinneyEtAl2008,Vergados2003,Widrow2005,Catena2012,Pato2013,FerrerHunter2013,Bhattacharjee2013,LavalleMagni2015,Boddy2017}, to its anisotropic extensions \cite{Osipkov1979,Merritt1985,Cuddeford1991,HunterEtAl1993,CuddefordLouis1995,UllioKamion2001,Wojtak2008,Hansen2009,Strigari2013,BozorgniaEtAl2013,FornasaGreen2014,Hunter2014,Cerdeno2016,Petac2018,Petac2019axisymmetric}---see ref.~\cite{Lacroix2018} for a critical review discussing theoretical issues---through more sophisticated methods such as the angle-action formalism \cite{McMillan2008,Posti2015,Williams2015,Binney2015,Sanders2016,Cole2017,Binney2020}. In this work, we focus on the Eddington inversion and its anisotropic extensions. The assumptions involved---especially spherical symmetry---are stronger than for angle-action methods, but even so the associated theoretical error on observables relevant to DM searches was shown to be only $\sim$~10-20\% based on careful comparisons with simulated galaxies \cite{Lacroix2020}. Moreover, with Eddington-like methods it is by construction straightforward to account for the correlations between the mass components of a galaxy and the resulting PSDF. In this work we argue that these approaches go a step further with respect to the ones discussed in the previous paragraphs, and provide a reliable next-to-minimal way to estimate astrophysical uncertainties on capture-related observables that self-consistently account for kinematic constraints. We demonstrate that this plays a very important part to harness the full potential of DM capture in stars as a probe of the properties of DM candidates, especially when considering very promising regions like the Galactic center (GC).

In Sec.~\ref{sec:capture} we outline the general formalism to compute the DM capture rate in stars, and in Sec.~\ref{sec:first_principles} we overview the Eddington inversion method and some of its anisotropic extensions, and their application to capture in stars in the MW. Then in Sec.~\ref{sec:results} we describe the systematic uncertainties on capture rates in different types of stars, from phase-space modeling. Finally, we summarize our main conclusions in Sec.~\ref{sec:conclusion}, to which we refer the reader interested in an overview of our results.

\section{Dark Matter capture}
\label{sec:capture}

\subsection{General formalism}
\label{ssec:capture_formalism}

Particles in the DM halo of the MW with a non-negligible scattering cross section can interact with the baryonic matter that composes stars. If the velocity of the DM particle after scattering is lower than the local escape velocity, it will get gravitationally bounded to the star. The capture rate per unit volume of a DM particle with mass $m_{\chi}$ is given by \citep{Gould1987}
\begin{align}
    \frac{\mathrm{d}C_i}{\mathrm{d}V}(r,r^\prime) = \frac{\rho_{\chi}(r)}{m_{\chi}} \int_0^{u_{\text{max},i}} \! \mathrm{d}u \, \frac{f_{u}^{*}(r,u)}{u} w \ \Omega_{v_{\text{esc}},i}^- \left( w\right)\,,
    \label{eq:cap_diff}
\end{align}
where $i$ labels the target nucleus, $\rho_{\chi}(r)$ is the DM density, $u = \left|\vec{u}\right|$ is the modulus of the DM particle velocity relative to the target star (at infinity), and $f_{u}^{*}(r,u)$ is the normalized speed DF of the DM halo in the star frame (away from the gravitational field of the star). Throughout this work we use $r^\prime$ to describe the radial position inside the star, as opposed to $r$, which is used to describe the radial position in the MW. The velocity at which the particle crosses the star shell at $r^\prime$ is given by 
\begin{align}
    w \equiv w(r^\prime) =\sqrt{u^2+v_{\text{esc}}(r^\prime)^2}\,,
\end{align}
where $v_{\text{esc}}(r^\prime) = \sqrt{2\left[\phi({\infty}) - \phi(r^\prime)\right]}$ is the escape velocity and $\phi(r^\prime)$ the gravitational potential of the star at the corresponding shell. The integral in Eq.~\eqref{eq:cap_diff} has an upper limit, $u_{\mathrm{max},i}$, which is the speed above which capture is kinematically forbidden through one single collision, discussed in Sec.~\ref{ssec:vel_space_capture}. Finally, $\Omega_{v_{\text{esc},i}}^- \left( w \right)$ is the rate at which a DM particle with velocity $w$ interacting with a nucleus $i$ will scatter to a velocity lower than the local escape velocity (the full expressions for the computation of the scattering rate $\Omega_{v_{\text{esc},i}}^- \left( w \right)$ are laid out in App.~\ref{appendixA}). While the DM-nucleon cross section enclosed in $\Omega_{v_{\text{esc},i}}^- \left( w \right)$ is model dependent, in this work, for the sake of discussion, we follow the standard practice in the literature and consider a non-relativistic energy- and velocity-independent coupling with constant cross section $\sigma_{\mathrm{const},i}$. Furthermore, we only consider interactions with nucleons, \ie, no electron scattering. As usual in the literature, we consider spin dependent (SD) and spin-independent (SI) contributions to the scattering cross section separately \cite{jungman1996},\footnote{\change{In this work we do not consider more general effective operators---which arise in effective field theories and introduce complex speed- and energy-dependent terms in the DM-nucleon differential scattering cross section \cite{Fitzpatrick2013}---or for scattering processed proceeding through light mediators \cite{Dasgupta2020}, for which the normalization of the capture rate can change appreciably. Nonetheless, although the interaction type can affect the velocity dependence of the integrand of the capture rate, its contribution is marginal when compared to the main velocity dependence that comes from the $f_{u}^{*}(u)/u$ term, so it does not affect our conclusions.}}
\begin{align}
    \sigma_{\mathrm{const},i}^{\mathrm{SI}} &= \sigma_{\mathrm{SI}} A_i^2 \frac{\left(m_{\chi}m_i\right)^2}{\left(m_{\chi}+m_i\right)^2} \label{eq:css1} \frac{\left(m_{\chi}+m_p\right)^2}{\left(m_{\chi}m_p\right)^2}\,, \\ \sigma_{\mathrm{const},i}^{\mathrm{SD}} &= \sigma_{\mathrm{SD}} \frac{4\left(J_i +1\right)}{3J_i} \left|\langle S_{\mathrm{p},i}\rangle + \langle S_{\mathrm{n},i} \rangle \right|^2\,,
    \label{eq:css2}
\end{align}
where $A_i$ is the atomic number of nucleus $i$, $J_i$ is the total spin of the nucleus and $\langle S_{\mathrm{p},i}\rangle$ and $\langle S_{\mathrm{n},i}\rangle$ are the spin expectation values of its proton and neutron systems. It should be noted that in the context of capture in stars, SD scatterings are highly suppressed for most elements, except for hydrogen, for which $\sigma_{\mathrm{const},i}^{\mathrm{SD}}=\sigma_{\mathrm{SD}}$. Finally, the total capture rate is simply given by
\begin{equation}
    C(r) = 4 \pi \int_0^{R_*}\! \sum_i \frac{\mathrm{d}C_i}{\mathrm{d}V}(r,r^\prime) {r^\prime}^2 \, \mathrm{d}r^\prime \,,
    \label{eq:total_cap}
\end{equation}
where the sum is performed over all target elements. 

Although the capture rate grows with the scattering cross section, it has a maximum value when the probability of a DM particle interacting with at least one nucleon is equal to one, \ie, when the star is completely optically thick. The critical cross section $\sigma_{\mathrm{crit}}$ that describes this geometric limit is defined by the relation
\begin{equation}
    \sum_i \sigma_{\mathrm{crit},i} N_i = \pi R_*^2\,,
\end{equation}
where $N_i$ is the number of nucleons for element $i$, $\sigma_{\mathrm{crit},i}$ is critical cross-section corresponding to element $i$ (see Eqs.~\ref{eq:css1} and \ref{eq:css2}), and the capture rate per unit volume in this limit is given by (\eg~\cite{Garani2019})
\begin{equation}
    C_{\mathrm{geo}}(r) = \pi R_*^2\ \frac{\rho_{\chi}(r)}{m_{\chi}} \int_0^{u_{i,\text{max}}} \mathrm{d}u \frac{f_{u}^{*}(r,u)}{u} w(R_*)^2 \,.
    \label{eq:cap_geo}
\end{equation}
The importance of the geometric limit is different for each type of star. For example, considering only SD interactions, for the Sun $\sigma_{\mathrm{crit}}$ is $ \approx 1.7 \times 10^{-35} \, \mathrm{cm}^2$, whereas for a typical NS with $M_* = 1 \, \rm M_{\odot}$ and $R_* = 10\, \mathrm{km}$ the critical cross section is $\sigma_{\mathrm{crit}} \approx 2.6 \times 10^{-45} \, \mathrm{cm}^2$. As a result, while for solar-like stars $\sigma_{\mathrm{crit}}$ sits orders of magnitude above current scattering cross section upper limits (approximately $10^{-45}-10^{-46} \ \mathrm{cm}^2$ for $m_{\chi} = 10-1000 \ \mathrm{GeV}$ \citep{pdg2018}), the geometric limit needs to be accounted for in the case of compact stars such as NSs.

It is important to note that while Eq.~\eqref{eq:cap_diff} is valid for capture through one single collision, it does not account for particles that end up captured after losing energy over multiple scatterings. While this is accurate for solar-like stars, where traversing DM particles will only interact at most once, the same is not true for compact stars, where DM particles can undergo multiple scatterings before leaving the interior of the star. In those cases, multi-scattering capture can be non-negligible, mainly for DM particles with $m_{\chi} \gtrsim 10 \ \mathrm{TeV}$ \citep{Dasgupta2019}. In the same way, effects due to the degeneracy of the medium in the star, which are also important in the case of compact objects such as NS stars, can lead to an important suppression of the capture rate for $m_{\chi} \lesssim 1 \ \mathrm{GeV}$ \cite{Garani2019}. However, we restrict our study to DM masses larger than 1~GeV to avoid having to model these effects, but our conclusions regarding the impact of phase-space modeling hold qualitatively for any DM mass. \change{We also note that in NSs a proper treatment of the relativistic kinematics of the target particles is in principle required \cite{Joglekar2020a,Joglekar2020b,Bell2020a,Bell2020b}. However, this mostly affects the interaction rate $\Omega_{v_{\text{esc}},i}^-$, while the dominant effect of the PSDF on the capture rate is encapsulated by the $f_{u}^{*}/u$ term, so in this work we still rely on the non-relativistic approximation. Moreover, for NSs we focus on the geometric limit, for which the actual kinematics do not play any role.}

\change{Finally, it should also be noted that in NSs, general relativistic (GR) effects become significant and modify Eqs.~\eqref{eq:cap_diff} and \eqref{eq:cap_geo} since DM particles can also be accelerated to relativistic velocities \cite{Bell2020a,Bell2020b}. By neglecting these corrections one underestimates the capture rate by about one order of magnitude for sub-GeV DM candidates \cite{Bell2020a,Bell2020b}. In the regime for which the geometric limit holds, \ie~for the DM mass range considered in this work, GR corrections boil down to a metric-dependent normalization factor (leading to a $\sim 70\%$ increase of the capture rate) and a slight decrease in the escape speed of the NS \cite{Bell2020a,Bell2020b}. While these effects are important for the normalization of the capture rate, in this work we are interested in the differences between the results obtained for different PSDFs. Given that the GR factors are essentially decoupled from the terms that depend on the PSDF, they cancel out when comparing the results for different PSDF models. The sole exception concerns the escape speed that is reduced in the GR framework and could change the velocity domain over which the PSDF is probed by capture. However, even if the GR correction is accounted for, the escape speed is still much larger than the most probable speed of DM particles, so it does not affect our estimates of the capture rate. The reason for this is that for NSs the speed integral in the capture rate (see Eq.~\ref{eq:cap_diff}) will still probe the entire speed DF, and only significant differences between speed DF models lead to sizable differences on the capture rate (see for instance the discussion in Sec.~\ref{ssec:vel_space_capture}).}

\subsection{Velocity distribution and capture rate: standard picture and limitations}

\change{As shown in Eq.~\eqref{eq:cap_diff}, the DM velocity DF is a central ingredient in the computation of the DM capture rate}. In particular, systematic uncertainties on the velocity DF will propagate to capture rates. This is true even in the case of a constant scattering cross section considered in this work. Here we recall the standard assumptions made in the literature and discuss their limitations.

Let $f_{\vec{v}}(\vec{r},\vec{v})$ be the velocity DF of DM particles in the Galactic frame at position $\vec{r}$.\footnote{The PSDF $f(\vec{r},\vec{v})$ and velocity DF $f_{\vec{v}}(\vec{r},\vec{v})$ are equivalent quantities that differ only by the fact that the PSDF is normalized to the total mass of the system, and the velocity DF to 1, and we have $f_{\vec{v}}(\vec{r},\vec{v}) = f(\vec{r},\vec{v})/\rho_{\chi}(\vec{r})$.} The usual approach used in the computation of the capture rate relies on the assumption that DM particles in the halo follow an isotropic MB velocity distribution. This gives in the Galactic frame
\begin{equation}
    f^{\rm MB}_{\vec{v}} (\vec{v}) = \frac{1}{\left(\pi v_{0}^2\right)^{3/2}} \exp\left(-\frac{\left|\vec{v}\right|^2}{v_{0}^2}\right)\,.
    \label{eq:shm}
\end{equation}
 The most probable speed $v_0$ is then generally fixed at the value of the circular speed at the Solar circle, $v_0 = v_{\odot} \approx 220\, \rm km\, s^{-1}$. It should be noted that this benchmark value is only valid in the Solar neighborhood. In the usual approach, the velocity dispersion $\sigma_{v}$ is related to the most probable speed $v_0$ via $\sigma_{v} = \sqrt{3/2} v_0$, giving $\sigma_{v} \approx 270\, \rm km\, s^{-1}$. In the following we refer to the MB velocity DF given in Eq.~\eqref{eq:shm}, with most probable speed $v_0 = v_{\odot} = 220\, \rm km\, s^{-1}$ everywhere in the Galaxy, as the standard halo model (SHM). This is the model most commonly used in the literature on capture-related DM searches, as analytic results and rough estimates of DM capture rates in stars can be readily obtained.\footnote{It should be noted that the MB distribution in Eq.~\eqref{eq:shm} must in fact be truncated at the Galactic escape speed $v_{\rm esc}^{\rm gal}$ at the Solar circle. However, in most cases that truncation is ad hoc and can introduce additional systematic uncertainties by changing the overall normalization of the distribution. Regardless, since analytic results on capture rates can only be obtained for the non-truncated distribution, the latter is used instead in most capture-related studies.}

However, a MB model for the speed DF is not predictive, and is bound to rely on ad hoc estimates of the peak speed $v_0$. In particular, the SHM only corresponds to a solution to the collisionless Boltzmann equation for an isothermal halo with $\rho(r) \propto r^{-2}$, which is expected to be approximately true only in the Solar neighborhood. Therefore, outside of this region the SHM is bound to be incompatible with the actual dynamics of the MW, and does not provide a satisfactory description of the DM phase-space \cite{King1966,BinneyEtAl2008,Kazantzidis2004,Sanchis2004,Wojtak2008}. This is especially problematic, since any phase-space model used in the context of velocity-dependent DM searches must account for kinematic constraints on the components of the Galaxy. One could argue that this could be solved by simply adopting a MB with a more realistic most probable speed $v_0$ that varies with radius and thus indirectly accounts for the dynamics of the Galaxy---\change{setting for instance $v_{0} = v_{\rm c}(r)$, where $v_{\rm c}(r)$ is the circular speed at radius $r$, or $v_{0} = \sqrt{2/3} \sigma_{v}^{\rm J}$ with $\sigma_{v}^{\rm J}$ the velocity dispersion found by solving the Jeans equation (\eg~\cite{BinneyEtAl2008}); see Sec.~\ref{ssec:MB_models}. While we still discuss these two cases in the following sections, their associated DFs are no longer a self-consistent solution to the collisionless Boltzmann equation. Therefore, even if they lead to correct order-of-magnitude estimates for capture observables (see also ref.~\cite{Lacroix2020})}, such an ad hoc adjustment of the velocity DF does not provide a full phase-space model. This is problematic since in that case the consistency of the velocity DF with the underlying Galactic mass model is hard to assess. Moreover, it is not possible to determine whether the model corresponds to a stable configuration of the DM halo or not. This makes it very difficult to evaluate the associated systematic uncertainties in a robust way. 

On the contrary, even if they are also based on simplifying assumptions, prediction methods relying on the Eddington inversion and its extensions do provide a theoretically grounded phase-space model that self-consistently accounts for kinematic constraints and predicts variations of the typical DM speed in the Galaxy, while making it possible to exclude pathological models. We will see in Sec.~\ref{sec:results} that this self-consistent approach leads to significant deviations on capture rates and subsequent observables with respect to the results obtained with the SHM---especially in the inner parts of the Galaxy---and allows for a robust estimate of systematic uncertainties based on physical arguments.

\subsection{Region of velocity space available for capture}
\label{ssec:vel_space_capture}

An important factor that will determine how the uncertainties on the velocity DF will impact on the DM capture rate is the velocity domain over which DM particles can be captured. For instance, direct detection experiments which attempt to observe nuclear recoils from DM interactions are only sensitive to the high velocity tail of the DM velocity distribution. Consequently, any uncertainty affecting this limited region of the velocity DF can have a dramatic effect on the prediction of the expected signal for direct DM searches in the low-mass region \cite{LavalleMagni2015}. Fortunately, the velocity-space region of interest for the computation of the capture rate encompasses a much more extensive fraction of the DM population in the Galactic halo. This somewhat mitigates the impact of uncertainties in specific regions of the velocity DF on the capture rate and associated observables. However, the effect is still significant and we quantify it in this work.

While there is no minimum velocity requirement for the capture of a given DM particle, there is a velocity $u_{\mathrm{max},i}$ above which it is impossible for a DM particle to lose the minimum energy required for capture in one collision. This means that DM capture is a priori sensitive to the low-velocity tail of the velocity DF. The maximum speed, which is a function of the target nucleus $i$, can be obtained through simple momentum and energy conservation considerations, which yield \citep{Choi2014}
\begin{align}
    u_{\mathrm{max},i}(r^\prime) =  \dfrac{2\sqrt{m_{\chi}m_i}}{\left| m_{\chi}-m_i \right|} v_{\mathrm{esc}}(r^\prime)\,.
    \label{eq:u_max}
\end{align}
It should be noted that $u_{\mathrm{max},i}$ is the maximum velocity for capture of a DM particle located at infinity, and not at the moment of collision with the target nuclei. As seen in Eq.~\eqref{eq:u_max}, the maximum velocity depends on the masses $m_{\chi}$ and $m_i$ of the DM and target nuclei, respectively. Then, on the one hand, a DM particle with $m_\chi = 5\, \mathrm{GeV}$ scattering off of a hydrogen nucleus on the surface of the Sun (where $v_{\mathrm{esc}} \approx 618\, \mathrm{km\, s^{-1}}$), will have a maximum velocity $u_{\rm max,H} \approx 691\, \mathrm{km\, s^{-1}}$, which is much larger than the most probable speed $v_0 = 220\, \mathrm{km\, s^{-1}}$ in the SHM. In this case, almost all DM particles from the Galactic halo can be captured in one single collision. On the other hand, if we consider DM particles with $m_\chi = 100\, \mathrm{GeV}$, only those with a velocity lower than $u_{\rm max,H} \approx 125\, \mathrm{km\, s^{-1}}$ are available for capture, which falls in the low-speed tail of the distribution. Assuming the SHM, this represents $\sim 10 \%$ of the DM population in the halo. In general terms, the bigger the difference between the mass of the target nucleus and $m_\chi$, and the smaller the escape speed from the star, the more susceptible the capture rate is to uncertainties on the low-velocity region of the PSDF. 

Keeping these caveats in mind, we perform a quantitative study of systematic uncertainties on the capture rate and related observables associated with Galactic DM phase-space modeling, using self-consistent Eddington-like prediction methods based on first principles. We discuss in particular these uncertainties as a function of the mass of the DM candidate, as well as the properties of the target stars.

\section{Speed distribution of dark matter particles from first principles}
\label{sec:first_principles}
\subsection{Overview}
\label{ssec:Eddington_overview}

In order to go beyond the simplistic approximation that DM particles follow a MB distribution with uniform velocity dispersion throughout the halo, it is necessary to take a closer look at the dynamical properties of the Galaxy, constrained by kinematic measurements. Equilibrium models based on the Eddington formalism and its extensions actually provide a simple but dynamically self-consistent picture of the phase space of DM particles in a galactic object like the MW, which accounts for observational constraints. 

In this section, we recall the formalism of equilibrium phase-space models from first principles. In Sec.~\ref{ssec:Eddington_inversion}, we briefly describe the standard Eddington inversion method, which relies in particular on the assumption of isotropy of the velocity tensor, and in Sec.~\ref{ssec:vdf_anisotropy} we describe the method we used to quantify the uncertainty on DM observables induced by the unknown degree of anisotropy of the velocity tensor, using an extension of the Eddington inversion.

For a system in dynamical equilibrium, \ie~for a virialized object, the PSDF $f(\Vec{r},\Vec{v})$ is solution to the collisionless Boltzmann equation and the Poisson equation, and by virtue of the Jeans theorem, the PSDF can be expressed as a function of integrals of motion \cite{Ollongren1962a,BinneyEtAl2008}. Under the assumption of spherical symmetry, one can write $f(\Vec{r},\Vec{v}) = F(\mathcal{E},L)$, where $L=|\vec{r}\times\vec{v}|$ is the modulus of
the angular momentum per unit mass, and $\mathcal{E} = \Psi(r) - v^{2}/2$ is the relative energy per unit mass, where $\Psi(r) = \Phi(R_{\mathrm{max}}) - \Phi(r)$ is the relative gravitational potential, with $\Phi(r)$ the total gravitational potential created by both DM and baryonic matter, and going to 0 at infinity, and $R_{\rm max}$ is a radius chosen to represent the boundary of the system. The relative potential can be expressed as
\begin{equation}
\Psi(r) = \int_{r}^{R_{\rm max}} \!  \mathrm{d}r' \, \dfrac{G m(r')}{r'^{2}} \,,
\label{eq:relative_potential}
\end{equation}
where the mass enclosed in a sphere of radius $r$ is
\begin{equation}
m(r) = 4 \pi \int_{0}^{r} \! \mathrm{d}r'\, r'^{2} \rho(r') \,.
\label{eq:mass}
\end{equation}

In this work, we focus on equilibrium models such as those provided by the Eddington inversion method, which are simple enough to take a given galactic mass model as input, while being fully self-consistent, which means that dynamical correlations within the object are correctly accounted for. However, these models can in principle strictly speaking only be applied to spherically symmetric systems, which in principle is not valid for a spiral galaxy like the MW which features prominent stellar and gas disks. Nevertheless, Eddington-like methods have the advantage over more sophisticated methods of making it straightforward to propagate uncertainties on the dynamics of the galactic object of interest all the way to quantities of interest for DM searches, such as capture rates in stars. Therefore, in this approach, in order to account for non-spherical components with profile $\rho(\vec{x})$---such as stellar and gas disks---we in fact consider the associated mass within radius $r$
\begin{equation}
\label{eq:mass_axisymm}
m(r) = \int_{|\vec{x}|\le r}\mathrm{d}^{3}\vec{x} \,\rho(\vec{x})\,,
\end{equation}
which allows us to compute a spherically symmetric approximation of the gravitational potential for the corresponding component. This ensures that our spherically symmetric models account for the non-spherical nature of the underlying mass as consistently as possible in the framework of Eddington-like methods. In practice, this provides a reasonable dynamical picture of a galactic object. Indeed, in ref.~\cite{Lacroix2020}, zoom-in simulations were used to test the Eddington method on MW analogues, and especially its relevance to the study of non spherically symmetric galactic systems. Extracting mass models of the simulated galaxies, sphericizing the non-spherically symmetric components, and feeding them to the Eddington inversion procedure, made it possible to predict the DM PSDF and some of its statistical moments that are relevant to DM searches. These predictions were then compared with the moments of the PSDF directly extracted from the simulation, and this led to the conclusion that the error on the moments that one makes by using the Eddington method is of the order of 10-20\%, all the way down to the resolution limit of the simulations, which is already deep in the region where non-spherical components are important. Therefore, provided the spherical symmetrization procedure is indeed applied to the non-spherical components, the presence of the latter does not prevent the Eddington method to already reach a very reasonable level of predictivity of 10-20\%, which is sufficiently accurate for the purpose of this work.

\subsection{Maximal symmetries: the Eddington inversion method}
\label{ssec:Eddington_inversion}

If, in addition to spherical symmetry, one assumes that the velocity tensor of the DM component is isotropic, the PSDF can be expressed as a function of the energy only: $f(\vec{r},\vec{v}) \equiv F(\mathcal{E})$. It is then possible to invert the relation between the PSDF and the DM mass density $\rho_{\chi}$,
\begin{equation}
\label{eq:rho}
\rho_{\chi}(r) = \int \! f(\mathcal{E})\, \mathrm{d}^{3}\vec{v}\,,
\end{equation}
and there is a one-to-one correspondence between the PSDF and a given density-potential pair, encapsulated in the well-known Eddington formula \cite{Eddington1916,BinneyEtAl2008}:
\begin{equation}
\label{Eddington_formula}
F(\mathcal{E}) = \dfrac{1}{\sqrt{8}\pi^{2}} \left[ \dfrac{1}{\sqrt{\mathcal{E}}} \left( \dfrac{\mathrm{d}\rho_{\chi}}{\mathrm{d}\Psi} \right)_{\Psi=0} + \int_{0}^{\mathcal{E}} \! \dfrac{\mathrm{d}^{2}\rho_{\chi}}{\mathrm{d}\Psi^{2}} \, \dfrac{\mathrm{d}\Psi}{\sqrt{\mathcal{E} - \Psi}}    \right]\, .   
\end{equation}
Here $\rho_{\chi}$ is the DM density, whereas $\Psi$ is the total (DM + baryons) gravitational potential. 

The divergence in phase space associated with the first term in Eq.~\eqref{Eddington_formula} is a spurious feature that arises when imposing a finite radial boundary to the system of interest, and translates into a divergence of the speed distribution close to the escape speed. This is especially important in the context of direct searches for low-mass DM. Self-consistent regularization procedures are discussed in ref.~\cite{Lacroix2018}. However, for DM capture the high-velocity tail does not play any role, and we can in practice disregard the diverging term, or equivalently consider that the system is infinite (\ie~let $R_{\rm max}$ go to infinity), which we do in the following.

\subsection{Maxwell-Boltzmann models beyond the SHM}
\label{ssec:MB_models}

\change{At first sight, the first step one could take to go beyond the simplest assumption of a VDF following a MB distribution with fixed $v_0 = v_{\odot} \approx 220\, \rm km\, s^{-1}$, in a way that accounts for variations of the characteristic speed of DM particles in the Galaxy, is to consider a MB model in which $v_0$ is given by the circular speed,}
\begin{equation}
v_{\rm c}(r) = \sqrt{\dfrac{Gm(r)}{r}}. 
\end{equation}
\change{This, however, is still a very simplistic model since it only accounts for DM particles on circular orbits. Nevertheless, there is a better way to get some physical insight into the PSDF of a gravitational system by considering the Jeans equation, which is obtained by taking the second velocity moment of the steady-state collisionless Boltzmann equation \cite{Jeans1915,Ollongren1962a,BinneyEtAl2008}. For an isotropic system, the velocity dispersion is $\sigma_{v}^{\rm J}(r) = \sqrt{\left\langle v^2 \right\rangle_{\rm J}(r)}$, where}
\begin{equation}
\left\langle v^2 \right\rangle_{\rm J}(r) = \frac{3\,G}{\rho}\int_{r}^{\infty}{\rm d}r'\,\frac{\rho(r')\,m(r')}{r'^2}\,.
\label{eq:velocity_dispersion_jeans}
\end{equation}
\change{Unlike the Eddington inversion, this approach does not rely on the full steady-state collisionless Boltzmann equation, so it is technically simpler, but at the expense of losing a significant amount of information on the PSDF, since it is forced to follow a MB distribution, and any non-Maxwellian features are not accounted for. In particular, although, by definition, the velocity dispersion $\sigma_{v}^{\rm J}(r)$ from the Jeans equation is identical to the one computed with the Eddington PSDF, other moments of the PSDF are by no means constrained to be equal for both models. This is especially relevant for DM capture which is essentially determined by the first inverse moment.} 

\change{Finally, it should be noted that even if they can in some cases be useful to get better estimates than the SHM, MB distributions with $v_0 = v_{\rm c}(r)$ or $v_0 = \sqrt{2/3} \sigma_{v}^{\rm J}(r)$ are ad hoc, since they are not by themselves solutions to the collisionless Boltzmann equation (unlike a MB distribution with uniform $v_0$).}

\subsection{Relaxing the assumption of isotropy}
\label{ssec:vdf_anisotropy}

As illustrated by numerical simulations, the approximation that the DM in galaxies has an isotropic velocity tensor is likely to break down at least in some regions of the DM halo, especially in the outskirts (\eg~\cite{Ludlow2011}). In that case the original Eddington PSDF is not strictly sufficient to account for the additional degrees of freedom involved, and the PSDF has to depend on the modulus of the angular momentum, $f(\Vec{r},\Vec{v}) = F(\mathcal{E},L)$ assuming spherical symmetry. A galactic component with an anisotropic velocity tensor can be characterized in terms of an anisotropy profile \cite{Binney1980},
\begin{equation}
\label{eq:beta}
\beta(r) = 1 - \frac{\sigma_{\theta}^2 + \sigma_{\phi}^2}{2\sigma_{r}^2}\,,
\end{equation}
where $\sigma_{r}$, $\sigma_{\theta}$, and $\sigma_{\phi}$ are the velocity dispersions in spherical coordinates. However, the anisotropy profile of the DM is unconstrained by observations, and reliable theoretical predictions for given galactic mass models are lacking as well, although some attempts exist \cite{Hansen2009,Hunter2014,Svensmark2019}. It might also be possible to jointly predict the PSDF of a galaxy and the anisotropy profile \cite{HunterEtAl1993,Petac2019axisymmetric}, but the accuracy of these predictions needs to be assessed carefully. At the same time, numerical simulations of galactic halos seem to favor anisotropy profiles essentially varying from close to 0 in the central regions to larger positive values in the outer parts, with however large halo-to-halo scatter \cite{Ludlow2011}. 

Therefore, in the absence of robust predictions, and in order to account for the uncertainty on the anisotropy of the DM velocity tensor, we follow the approach of ref.~\cite{Wojtak2008}, in which the authors proposed an ansatz for the PSDF associated with an anisotropy profile defined by three parameters, namely an asymptotic value $\beta_{0}$ at the center, another one in the outskirts of the galaxy, $\beta_{\infty}$, and a characteristic angular momentum $L_{0}$ that sets the transition radius between both regimes. The ansatz assumes separability in terms of energy and modulus of the angular momentum, and reads
\begin{equation}
\label{eq:f_anis}
F(\mathcal{E},L) = f_{\mathcal{E}}(\mathcal{E}) \left( 1 + \dfrac{L^2}{2L_{0}^2} \right)^{-\beta_{\infty}+\beta_{0}} L^{-2\beta_0}\,.
\end{equation}
Although this PSDF model is unlikely to encapsulate all the complexity of the dynamics of the actual MW---in particular it assumes spherical symmetry---it was found to reproduce rather well the PSDF and anisotropy profile of simulated galaxy clusters in ref.~\cite{Wojtak2008}. Therefore, although detailed tests are needed in galactic structures, this ansatz still provides a realistic enough toy model for an anisotropic system, and as such represents an improvement over the standard Eddington solution. This functional form is a generalization of the constant-anisotropy model \cite{Henon1973,KentGunn1982,BinneyEtAl2008} and has more flexibility to account for a wider range of anisotropic models observed for instance in synthetic galactic halos \cite{Ludlow2011}. It is then possible to invert the integral relation between the DM density and PSDF,
\begin{equation}
\label{eq:rho_anis}
\rho_{\chi}(r) = \int \! F(\mathcal{E},L)\, \mathrm{d}^{3}\vec{v}\,.
\end{equation}
As discussed in App.~B of ref.~\cite{Wojtak2008}, Eq.~\eqref{eq:rho_anis} can be rewritten as an integral over the relative energy, similar to the one used to perform the standard Eddington inversion:
\begin{equation}
\label{eq:rho_anis_for_inversion}
\rho_{\chi}(r) = (2\pi)^{3/2}2^{-\beta_{0}}
\frac{\Gamma(1-\beta_{0})}{\Gamma(3/2-\beta_{0})} r^{-2\beta_{0}}
\int_{0}^{\Psi(r)}f_{\mathcal{E}}(\mathcal{E})(\Psi(r)-\mathcal{E})^{1/2-\beta_{0}}K(\Psi(r),\mathcal{E})\mathrm{d}\mathcal{E}\,,
\end{equation}
where $\Gamma$ is the Gamma function and
\begin{equation}\label{kernel}
K(\Psi(r),\mathcal{E})=(1+x)^{-\beta_{\infty}+\beta_{0}}
{}_{2}F_{1}(1/2,\beta_{\infty}-\beta_{0},3/2-\beta_{0},x/(1+x))\,,
\end{equation}
with $x \equiv r^{2}(\Psi(r)-\mathcal{E})/L_{0}^{2}$ and ${}_{2}F_{1}$ is the hypergeometric function. Unlike in the isotropic case, in order to determine $f_{\mathcal{E}}(\mathcal{E})$, in the most general case the inversion of the Volterra equation in Eq.~\eqref{eq:rho_anis_for_inversion} has to be performed numerically. In this work we used the algorithm presented in App.~B of ref.~\cite{Wojtak2008}. We do not repeat the description of the procedure here but instead refer the reader to the extensive discussion in the original reference. We checked our results by reconstructing the density following Eq.~\eqref{eq:rho_anis_for_inversion}---which serves as a test of consistency of the PSDF with the mass model---and by verifying that the velocity DFs we obtained were properly normalized. We also recover the PSDFs corresponding to the constant anisotropy model, for which the inversion is semi-analytic. Finally, our results for the velocity DFs are consistent with the results of ref.~\cite{FornasaGreen2014}.

\subsection{Milky Way mass models}
\label{ssec:mass_models}

In this study, to build our phase-space models, we rely on mass models for the MW developed in ref.~\cite{McMillan2017}, which we refer to as the McM17 mass models in the following. In that paper, the author accounted for a variety of kinematic measurements at $\sim$kpc scales such as proper motions of masers and the vertical motion of the Sun, and derived best-fit models for the density profiles of the various components of the MW, namely a stellar bulge, two stellar disks, two gas disks and a DM halo. The best fits of ref.~\cite{McMillan2017} were obtained for fixed values of the inner slope $\gamma$ of the DM profile, since the latter could only be poorly constrained. In this work, in order to bracket the uncertainty on the inner slope, we consider two cases, namely $\gamma = 1$, corresponding to the Navarro-Frenk-White (NFW) profile, and $\gamma = 0.25$, which is one of the cases considered in ref.~\cite{McMillan2017}, and representative of a cored halo. In particular, the case of a cored profile is important as it seems to be favored by a detailed analysis of the dynamics of the Galactic bulge and bar \cite{Portail2017}. The results we obtain throughout this work with these two characteristic generalized NFW profiles are also representative of alternative profiles such as the Einasto profile, which can go from cuspy to cored depending on the value of the logarithmic slope parameter. The density profiles of the various components are described in App.~\ref{appendixB}.\footnote{Here we do not consider $\gamma = 0$ since, as discussed in ref.~\cite{Lacroix2018}, a combination of such a DM profile with a baryonic components results in an Eddington PSDF which is likely to be an unstable solution of the collisionless Boltzmann equation. Therefore the Eddington formalism is not appropriate in that case. However, $\gamma = 0.25$ is actually representative of a cored profile, and provides in most cases \textit{a priori} physical solutions.}

We stress that the McM17 mass models rely on pre-\textit{Gaia} data, therefore unlike more recent studies (\eg~\cite{Cautun2020}), they do not account for the more recent data from the second \textit{Gaia} data release. However, our aim is to provide a reliable, although simple, phase-space model that goes a step beyond the standard approach, and that allows us to discuss the theoretical uncertainties related to the PSDF itself. Moreover, the dynamical picture has not changed significantly, even quantitatively, and the new mass models derived from \textit{Gaia} data for instance in ref.~\cite{Cautun2020} are close to the McM17 ones, and the DM and stellar parameters from ref.~\cite{McMillan2017} and ref.~\cite{Cautun2020} are consistent at the 68\% level. In practice, the most striking difference between the older and more recent mass models turns out to be a preference for a less massive DM halo in the \textit{Gaia} results. For instance the scale radius of an NFW halo and associated density at the Sun are found to be $r_{\rm s} = 14.4^{+4.5}_{-3.5}\, \rm kpc$, $\rho_{\chi,\odot} = 0.33 \pm 0.02\, \rm GeV\, cm^{-3}$ \cite{Cautun2020}, to be compared with $r_{\rm s} = 18.6^{+5.3}_{-4.4}\, \rm kpc$ and $\rho_{\chi,\odot} = 0.38 \pm 0.04\, \rm GeV\, cm^{-3}$ (McM17) in ref.~\cite{McMillan2017}. Although these values are compatible at 68\% confidence level, the smaller DM mass found in the latest results leads to a smaller typical DM speed in the MW. Therefore, with a mass model accounting for the more recent data the capture rate should be slightly larger than we find here (see for instance Sec.~\ref{ssec:capture_sun_like}) and further away from the SHM. Nevertheless, we argue that the McM17 mass models do allow us to provide a reasonable estimate of the capture rate that self-consistently accounts for the underlying dynamics of the Galaxy, and provide a test-bed to quantify the uncertainties on DM capture associated with the PSDF model. We insist that one of the main goals of our work was to draw the attention of the DM-capture community to the issue of phase-space modeling, and provide a simple model that encapsulates the main-order corrections on capture-related observables with respect to the standard approach relying on the SHM with uniform velocity dispersion.

An additional motivation for relying on the McM17 models is that these are complete mass models derived for several values of the inner slope $\gamma$ of the DM density profile, which turns out to be one of the strongest sources of systematic uncertainties on capture rates, with dramatic differences between cuspy and cored profiles. This allows us to quantify the uncertainty from both the DM phase-space model and density profile in a self-consistent way. 

There are also systematic uncertainties associated with the mass models themselves, not captured by the best-fit values of the parameters. To estimate them consistently, it is necessary to use a full Monte Carlo analysis of the kinematic data. Such an analysis, in particular on \textit{Gaia} data, is beyond the scope of the present work. Nevertheless this is not crucial since the corresponding systematic uncertainties, although sizable, are in any event subdominant with respect to errors from the inner slope of the DM profile and the properties of the PSDF (\eg~the level of anisotropy, as discussed in ref.~\cite{FornasaGreen2014}).

Finally, it should be noted that the McM17 mass models are not designed to accurately describe the dynamics at sub-kpc scales, since they are constrained by kpc-scale observations. As a result, in order to be able to model the DM phase-space distribution down to scales of a few pc, we complement the McM17 mass models by the model developed in ref.~\cite{Sofue2013} for the pc-scale nuclear star cluster---referred to as the inner stellar bulge in ref.~\cite{Sofue2013}---obtained by fitting rotation curve data.\footnote{We do not use the other components from ref.~\cite{Sofue2013}, since in that study the fits are only based on rotation curve data, so the stellar bulge, discs and DM halo are likely to be better constrained in ref.~\cite{McMillan2017}, although at larger radii.} This ensures that our phase-space models do account for observational constraints both in the inner region of the MW, and at larger scales, in particular at the position of the Sun. This also makes it possible to self-consistently quantify the uncertainties on capture-related observables from the unknown inner slope of the DM profile, the DM anisotropy profile, and to quantify the error made by using the SHM with a fixed velocity dispersion throughout the Galaxy.

\subsection{Estimating the uncertainty on the speed DF from the unknown anisotropy}
\label{ssec:uncertainty_anisotropy}

The functional form in Eq.~\eqref{eq:f_anis} allows us to marginalize over the three parameters that characterize the anisotropy. We define the ranges for these parameters as follows, in the line of the discussion in ref.~\cite{FornasaGreen2014}. As discussed in ref.~\cite{Lacroix2018} and references therein, some parameter configurations can lead to potentially unstable solutions to the collisionless Boltzmann equation when using Eddington-like inversion methods. However, models for which the energy part of the PSDF is a monotonically increasing function of the energy do provide stable solutions (\eg~\cite{Doremus1973}). Although this is only a necessary conditions, this allows us to filter out pathological models. Therefore, in this work we use this stability condition to retain only sound models, which in turn excludes portions of the \textit{a priori} parameter space for anisotropic parameters.

In order for the PSDF to be non-negative, we must have $\beta_0 \leq \gamma/2$ (cusp slope-central anisotropy relation), where $\gamma$ is the slope of the DM profile in the inner region \cite{AnEvans2006}. In principle this gives $\beta_0 \leq 0.5$ for NFW and $\beta_0 \lesssim 0.1$ for $\gamma = 0.25$. However, simulations tend not to find radial anisotropy in the inner regions of halos (\eg~\cite{Ludlow2011}), and in practice we find that PSDFs with $\beta_0 > 0$ do not satisfy the necessary condition for stability.\footnote{This was not accounted for in ref.~\cite{FornasaGreen2014} but as discussed extensively in ref.~\cite{Lacroix2018}, it is not satisfactory to estimate astrophysical uncertainties by considering PSDFs that are potentially unstable solutions of the collisionless Boltzmann equation.} Therefore, a more sensible upper limit is given by $\beta_0 \leq 0$. We also impose a lower bound on the central anisotropy, $\beta_0 \geq -0.5$, based on the results of cosmological simulations which tend to find values of the central anisotropy that are not significantly negative (\eg~\cite{Ludlow2011}).

There are a priori no theoretical constraints on $\beta_{\infty}$, but in practice the value of the anisotropy in the outskirts of galactic halos in cosmological simulations do not go above 1 \cite{Ludlow2011}. Moreover, as we do for the central anisotropy, we take $\beta_{\infty} \geq -0.5$, and we also allow for $\beta_{\infty} < \beta_{0}$ to give more flexibility to the model to account for a possible decrease in $\beta(r)$ above $\sim 5r_{\rm s}$ as observed in simulated galactic halos \cite{Ludlow2011}.

Finally, the parameter $L_0$ that sets the transition between the two different regimes in this anisotropic model can be expressed in units of the scale angular momentum $L_{\rm s} = r_{\rm s} v_{\rm s}$, where $r_{\rm s}$ is the scale radius of the DM profile and $v_{\rm s} = \sqrt{4 \pi G \rho_{\rm s} r_{\rm s}^2}$. $L_0 \sim L_{\rm s}$ corresponds to a transition radius $\sim 4 r_{\rm s}$. Here we consider $10^{-1} \leq L_0/L_{\rm s} \leq 10$. For values of $L_0$ typically larger than $10 L_{\rm s}$, the regime in which the anisotropy is equal to $\beta_0$ extends way above $r_{\rm s}$, essentially throughout the entire galaxy. In the opposite regime of small $L_0$, we do not go below $10^{-1} L_{\rm s}$ since in that case we are in the regime where $\beta_{\infty}$ actually represents the central anisotropy, and we end up in a situation where the cusp slope-central anisotropy relation is in most cases not satisfied, leading to non physical PSDFs.

After performing a Galilean boost to the stellar frame as described in Sec.~\ref{ssec:VDF_stellar_frame}, we obtain the speed DF marginalized over the anisotropy parameters sampled over the ranges of $\beta_0$, $\beta_{\infty}$ \change{(uniformly in linear space)} and $L_0$ \change{(uniformly in log space).} \change{This choice of priors was already made in ref.~\cite{FornasaGreen2014} and corresponds to uninformative priors (within the ranges defined above\footnote{It turns out that for the cored profile with $\gamma = 0.25$, some values of the anisotropy parameters lead to PSDFs that do not satisfy the necessary conditions for stability. As a result we exclude the corresponding values from our estimate of the marginalized speed DF. More specifically we exclude $\beta_{\infty} \geq 0.5$ for $L_{0} \leq 0.1$.}) since the anisotropy parameters are essentially unconstrained. Finally we compute the median speed DF and the corresponding 68\% and 95\% confidence regions.}

\subsection{Speed distribution in the stellar frame}
\label{ssec:VDF_stellar_frame}

 For a target star with velocity $\vec{v}_*$, the velocity DF $f^*_{\vec{u}}(r,\vec{u})$ in the star frame is computed from the velocity DF $f_{\vec{v}}$ in the Galactic frame through a Galilean boost via
\begin{equation}
    f^*_{\vec{u}}(r,\vec{u}) = f_{\vec{v}} \left(r, \vec{u} + \vec{v}_* \right)\,.
    \label{eq:vdf_boost}
\end{equation}
The speed DF---distribution of the modulus of the velocity---can then be obtained by integrating the velocity DF over the solid angle,
\begin{equation}
    f^*_u(r,u) = \left|\vec{u}\right|^2 \int \! f^*_{\vec{u}}(r,\vec{u})\, \mathrm{d} \Omega\,.
    \label{eq:vdf_mod}
\end{equation}
For a MB distribution, the integral over solid angle in Eq.~\eqref{eq:vdf_mod} can be solved analytically, by using $\left|\vec{v}\right|^2 = \left|\vec{u}\right|^2 + \left|\vec{v}_*\right|^2 + 2 \cos \theta \left|\vec{u}\right| \left|\vec{v}_*\right|$ and integrating over $\theta$, yielding
\begin{equation}
\label{eq:fu_mb}
f^{*}_{u, \rm MB}(u;v_{0}) = \frac{u}{\sqrt{\pi} v_* v_0} \left[\exp\left(-\frac{\left(u-v_*\right)^2}{v_{0}^2}\right) - \exp\left(-\frac{\left(u+v_*\right)^2}{v_{0}^2}\right)\right]\,,
\end{equation}
where $u \equiv \left|\vec{u}\right|$ and $v_* \equiv \left|\vec{v}_*\right|$. \change{The dependence on $r$ is in general omitted in the literature on capture, which relies on the SHM for which $v_0$ is generally fixed at $v_\odot$ everywhere in the Galaxy, so $f_{\rm SHM}(u) = f^{*}_{u, \rm MB}(u;v_\odot)$. As we argue in this paper, this is in most cases an overly simplistic approximation. Within the MB approximation, one can actually go further and start accounting for the dynamics of the Galaxy by setting $v_{0}$ to the circular speed $v_{\rm c}(r)$ of DM particles at the radius $r$ of interest, or by taking $v_{0} = \sqrt{2/3} \sigma_{v}^{\rm J}(r)$. The latter approach has the merit of relying on first principles through the Jeans equation, although a significant amount of information is lost by only considering the second moment of the collisionless Boltzmann equation. As we will discuss below, although these more evolved MB models can in some cases give relatively good approximations to the results obtained with more realistic models, they are by definition unable to account for non-Maxwellian features, unlike Eddington-like methods, and can therefore bias subsequent predictions.}

For an isotropic system with a PSDF $F(\mathcal{E})$ given by Eddington's inversion, the normalized speed distribution in the frame of the star reads
\begin{equation}
f^*_u(r,u) = 2 \pi u^2 \int_{0}^{\pi} \! \mathrm{d}\theta \, \sin \theta \, \dfrac{F(\mathcal{E})}{\rho_{\chi}(r)}\,,
\end{equation}
where $\mathcal{E} = \Psi(r) - \frac{1}{2}\left(u^2 + v_*^2 + 2 u v_* \cos \theta \right)$, while for an anisotropic system, the solid angle integral can be computed via (e.g.~\cite{UllioKamion2001})
\begin{equation}
f^*_u(r,u) = u^2 \int_{0}^{2\pi} \! \mathrm{d}\psi \, \int_{0}^{\pi} \! \mathrm{d}\eta \, \sin \eta \, \dfrac{F(\mathcal{E},L)}{\rho_{\chi}(r)}\,,
\end{equation}
where 
\begin{equation}
\mathcal{E} = \Psi(r) - \frac{1}{2}\left(u^2 + 2 u v_* \sin \psi \sin \eta + v_*^2 \right)\,,
\end{equation}
and
\begin{equation}
L = r \left(u^2 \sin^2 \eta + 2 u v_* \sin \psi \sin \eta + v_*^2\right)^{1/2}\,.
\end{equation}
\change{For $v_*$---the speed of the target star in the Galactic frame---the most common assumption in the literature is also $v_* = v_\odot$, even when the target is distinct from the Sun. So throughout this work, when considering the SHM we use $v_* = v_\odot$. However, for all the other models we take a more realistic estimate, $v_* = v_{\rm c}(r)$, where the circular speed $v_{\rm c}(r)$ is computed from the same mass model we use to derive the PSDF.} It should be noted however that in practice non-circular motion can amount to $\sim \pm 20-30\%$ of the circular speed, as noted for instance for molecular gas in ref.~\cite{Sofue2013}. We estimated that this induces an additional systematic uncertainty of at most 40\% on capture rates in the inner parts of the MW that should be taken into account in DM capture studies.

\subsubsection{Variation of the speed distribution in the Galaxy}
\label{sssec:variation_vdf}

\begin{figure}[t!]
\centering
\includegraphics[width=0.312\linewidth]{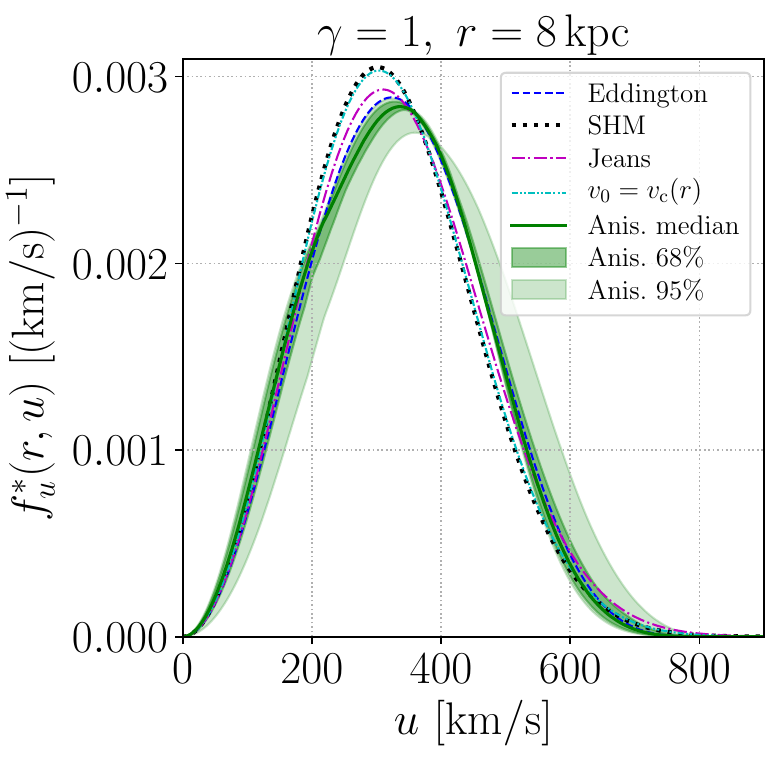} \hfill
\includegraphics[width=0.325\linewidth]{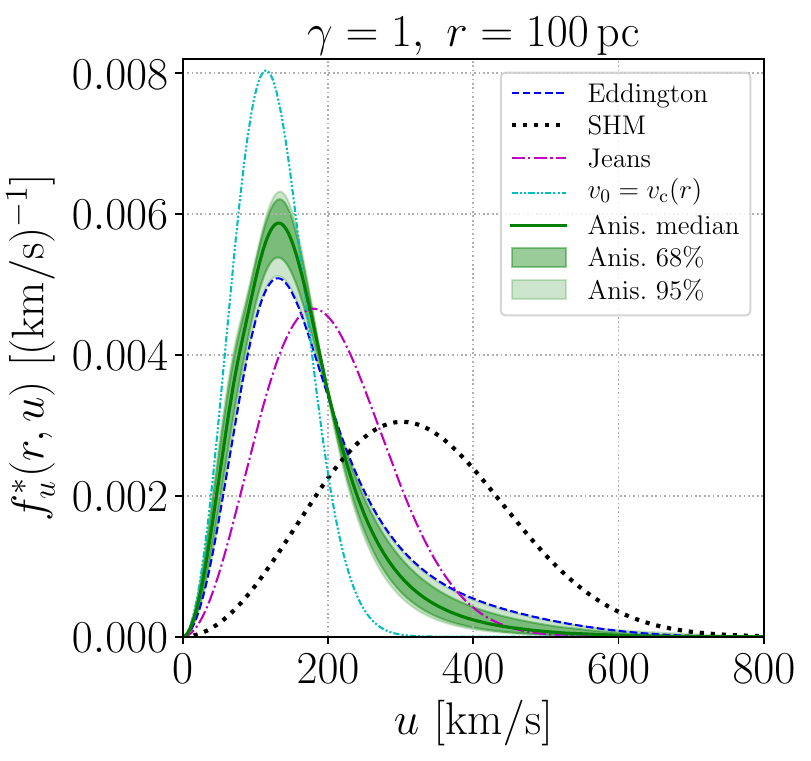} \hfill
\includegraphics[width=0.325\linewidth]{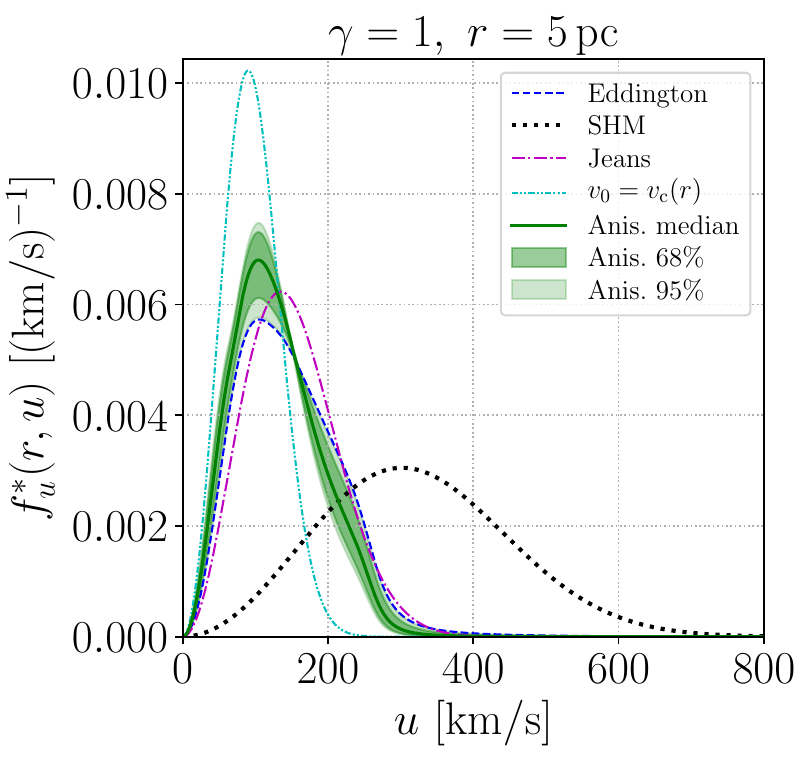}

\includegraphics[width=0.312\columnwidth]{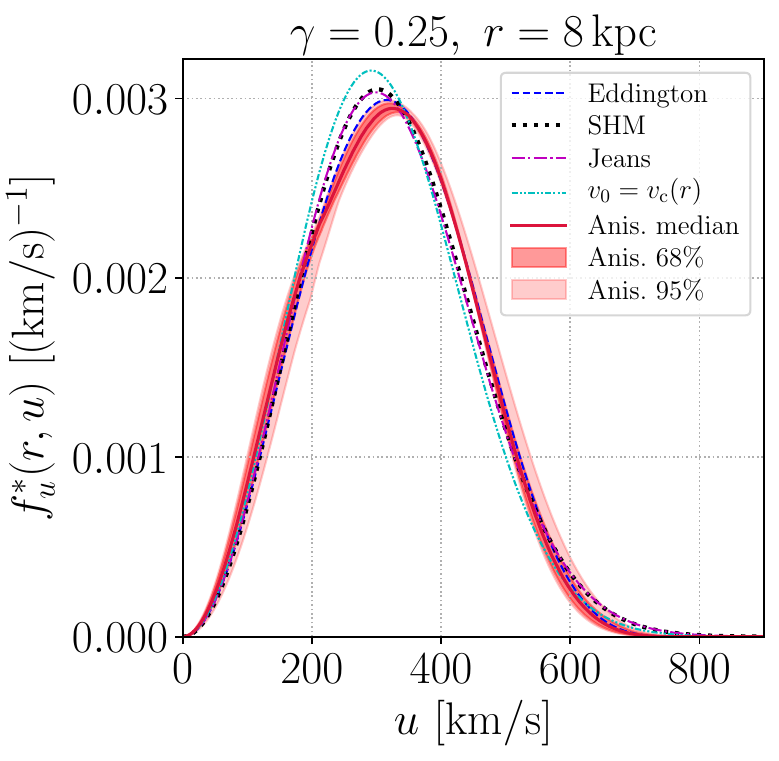} \hfill
\includegraphics[width=0.325\columnwidth]{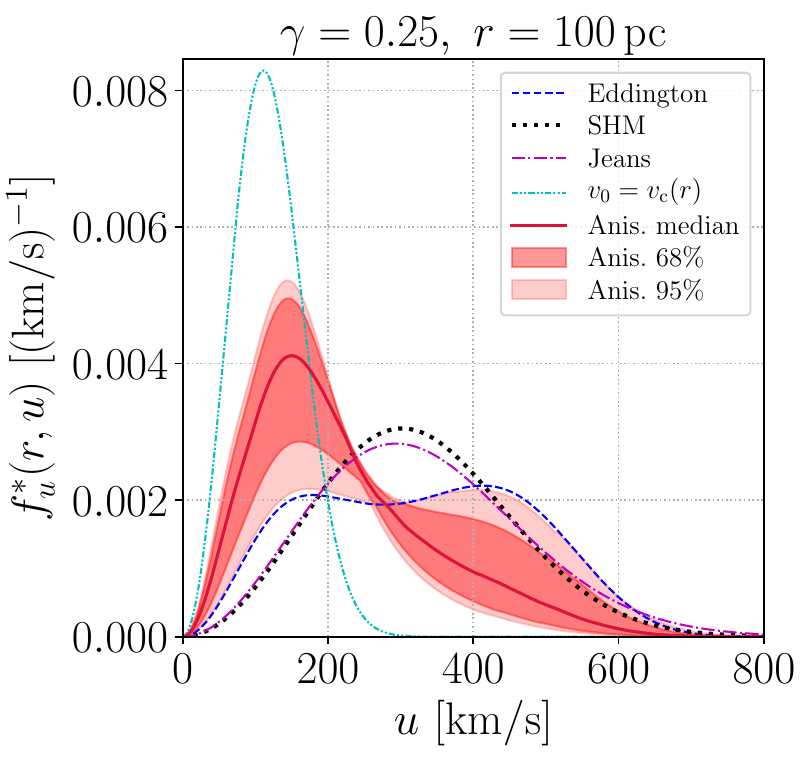} \hfill
\includegraphics[width=0.325\columnwidth]{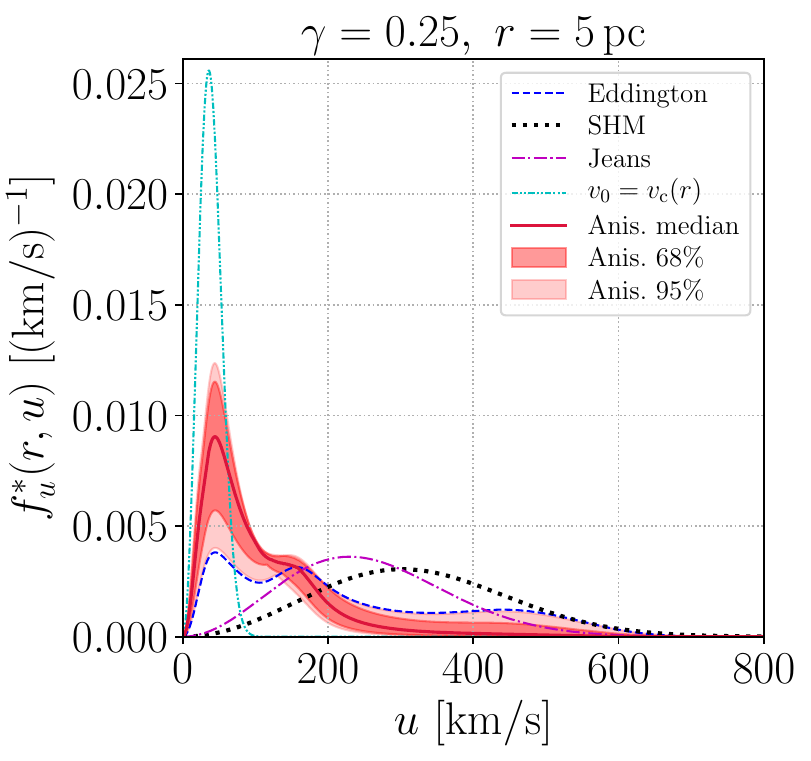}
\caption{Speed DF in the frame of stars located at three galactocentric radii: 8 kpc (left panels), 100 pc (middle panels) and 5 pc (right panels), for Galactic mass models associated with a DM profile of slope $\gamma = 1$ (upper panels, green) and $\gamma = 0.25$ (lower panels, red). Solid lines correspond to the median of the speed DFs obtained by sampling over the various anisotropy parameters. The dark and light shaded bands represent the associated 68\% and 95\% containment regions, which quantify the uncertainty from the unknown anisotropy. The results of the Eddington inversion that assumes isotropy are also shown as dot-dashed blue lines. \change{The speed DFs for MB models with $v_0 = v_{\odot} = 220\ \rm km\, s^{-1}$ (i.e.~the SHM), $v_0 = v_{\rm c}(r)$, and $v_{0} = \sqrt{2/3} \sigma_{v}^{\rm J}(r)$ (labeled `Jeans'), are included in all panels as black dotted, cyan dot-dot-dashed, and magenta dot-dashed lines, respectively. For the SHM the speed of the star is fixed at $v_* = v_\odot$, while for all the other models we take $v_*(r) = v_{\rm c}(r)$.}}
\label{fig:fu_boosted}
\end{figure}

The potential of capture in stars as a DM probe is actually not limited to the Sun or NSs in the Solar neighborhood. Although observations are significantly more challenging when moving away from the Sun, high DM-density regions such as the GC turn out to be even more promising to probe DM capture, especially in compact stars (\eg~\cite{Garani2019}). In order to assess systematic uncertainties from phase-space modeling on capture rates and subsequent observables towards the GC---see Sec.~\ref{sec:results}---we now discuss the evolution of the DM speed DF, modeled in a self-consistent way, as a function of radius. The speed DFs from the Eddington method and the anisotropic model in the frame of stars located at three benchmark values of the galactocentric radius $r$, namely 8 kpc---representative of the Solar neighborhood, 100~pc, and 5~pc, are shown in Fig.~\ref{fig:fu_boosted} in the left, middle, and right panels, respectively, for $\gamma = 1$ and $\gamma = 0.25$ in the upper and lower rows, respectively.\footnote{It should be noted that we do not consider radii smaller than 5~pc in order to avoid having to account for the effect of the central supermassive black hole Sgr A* which starts to dominate the gravitational potential below that radius.} The Eddington prediction (blue dot-dashed) is different from the median of the distribution of anisotropic speed DFs (solid lines, green for $\gamma = 1$ and red for $\gamma = 0.25$) because we do not consider radial anisotropy in the central region of the DM halo but we only sample negative values of the anisotropy parameter $\beta_0$. Therefore in that region the range of values of the anisotropy is not symmetric around 0. The difference is smaller in the outer part of the halo---above kpc scales---since in that region we probe the outer part of the anisotropy profile associated with $\beta_{\infty}$, for which we consider a range that is essentially symmetric around 0. The difference is also smaller for $\gamma = 1$ compared with $\gamma = 0.25$, since anisotropic speed DFs are actually sensitive to a modified density profile---in the inner region rescaled by $r^{2\beta_0}$ as shown in Eq.~\eqref{eq:rho_anis_for_inversion}---and the relative difference between the slope $2\beta_0 - \gamma$ of this effective profile and the slope $\gamma$ of the initial one is for most anisotropy parameters much larger for $\gamma = 0.25$ than for $\gamma = 1$.\footnote{For $\gamma = 0.25$ the speed distribution has an additional double- or triple-bump structure. This is only present for isotropic or radially anisotropic models, and is related to the effect of baryons that are much more prominent in the inner halo for a cored DM profile. Although this might signal some stability issues, the associated PSDF still satisfies the necessary condition for stability, so here we remain agnostic about the origin of these multiple populations in the speed DF.}

To illustrate the difference between the \change{MB models and the more realistic Eddington-like ones, we also include in each panel of Fig.~\ref{fig:fu_boosted} MB distributions in the stellar frame (given in Eq.~\eqref{eq:fu_mb}) with $v_0 = 220\, \rm km\, s^{-1}$, $v_0 = v_{\rm c}(r)$, and $v_0 = \sqrt{2/3} \sigma_{v}^{\rm J}(r)$.} 
The systematic uncertainty from the unknown anisotropy is illustrated by the 68\% and 95\% credible regions shown in each panel as dark and light shaded bands, respectively.\footnote{The 68\% and 95\% uncertainty bands are wider for $\gamma = 0.25$ in the inner region again because of the underlying effective DM profile involved.} Quantitatively, for capture-related observables, which in most cases probe the low-velocity tail of the speed DF, Fig.~\ref{fig:fu_boosted} shows that while \change{at 8~kpc and for low speeds} the speed DF for the SHM lies within the 68\% credible region \change{of} the anisotropic distributions, it underestimates the self-consistent results by a factor $\sim 10$ at 100~pc and 5~pc as evidenced by the values of the speed DFs at $50\, \rm km\,s^{-1}$, representative of the low-velocity tail. This translates into similar discrepancies in the capture rate, as discussed in Sec.~\ref{ssec:capture_sun_like}. 


\change{The Jeans-based MB speed DF, with $v_0 = \sqrt{2/3} \sigma_{v}^{\rm J}(r)$, lies much closer to the Eddington speed DF for the $\gamma = 1$ case than for the cored profile with $\gamma = 0.25$. In the latter case, the gravitational potential created by baryons plays a more important role in shaping the VDF, leading to more significant departures from a Gaussian distribution as compared with the cuspy case. This means that once the velocity dispersion of the speed is properly accounted for (as is the case with the Jeans model), the VDF for $\gamma = 1$ is close enough to a MB distribution so that the Jeans model provides a reasonable approximation to the Eddington prediction, although it still underestimates the low-velocity tail at small velocities. For the $\gamma = 0.25$ case, the VDF deviates more significantly from a MB distribution due to baryons, so that although the Jeans model gives the right velocity dispersion, it gives a speed DF close to the SHM, with significant discrepancies with respect to the Eddington-like models. Finally, the MB model based on the circular speed, $v_0 = v_{\rm c}(r)$ predicts roughly the right position of the most prominent peak of the speed DF, but it actually significantly overestimates the low-velocity tail. As a result, even if imposing that the velocity dispersion of the MB distribution is determined by the circular speed may appear a priori as an improvement over the SHM, it turns out that this significantly biases the speed DF towards low velocities---by ignoring more excentric orbits for which the velocity can get larger than the circular velocity close to pericenter passage---and ends up overtestimating capture, as shown in Sec.~\ref{ssec:capture_sun_like}. As a result, one must be careful when dealing with MB models.}

This \change{discussion} illustrates how modeling carefully the PSDF of DM particles becomes more and more important when moving towards the Galactic center\change{---where the VDF is more sensitive to the baryonic gravitational potential and therefore deviates more significantly from a Gaussian distribution. Moreover, this shows} how the extension of the Eddington inversion method to a DM halo with non-zero velocity anisotropy allows us to quantify the systematic uncertainty on the speed distribution, and subsequently the capture rate, in a self-consistent way.

An additional important factor to take into account when estimating these systematic uncertainties is to make sure that the phase-space models involved are physical. As discussed in Sec.~\ref{ssec:uncertainty_anisotropy}, stability criteria significantly restrict the parameter space of anisotropic models that give viable solutions to the collisionless Boltzmann equation. More specifically, for the NFW profile, if we allow for $\beta_0 > 0$ at small radii (\eg~5~pc) the systematic error bands shown in the upper left panel of Fig.~\ref{fig:fu_boosted} would be approximately 5 times wider than the one we obtain when including only physical solutions. This demonstrates the importance of relying on full phase-space models, like those provided by Eddington-like methods, \change{as opposed to ad hoc MB models with a velocity dispersion varying as a function of radius such as those discussed above.}

\subsubsection{Dependence on the maximum velocity for capture in one collision}
\label{sssec:maxvel_dep}

\begin{figure}[!t]
    \centering
    \includegraphics[width=0.49\columnwidth]{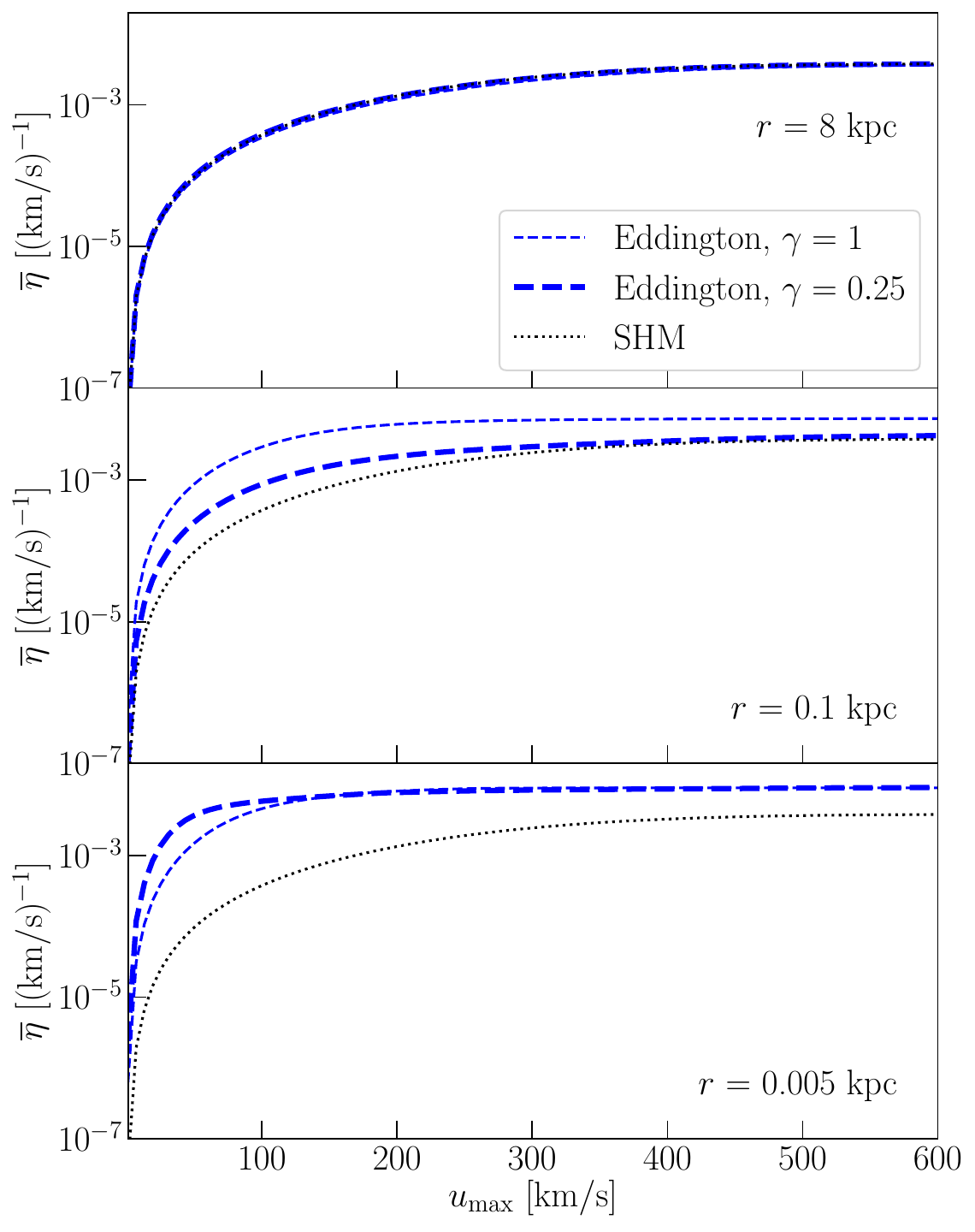} \hfill \includegraphics[width=0.49\columnwidth]{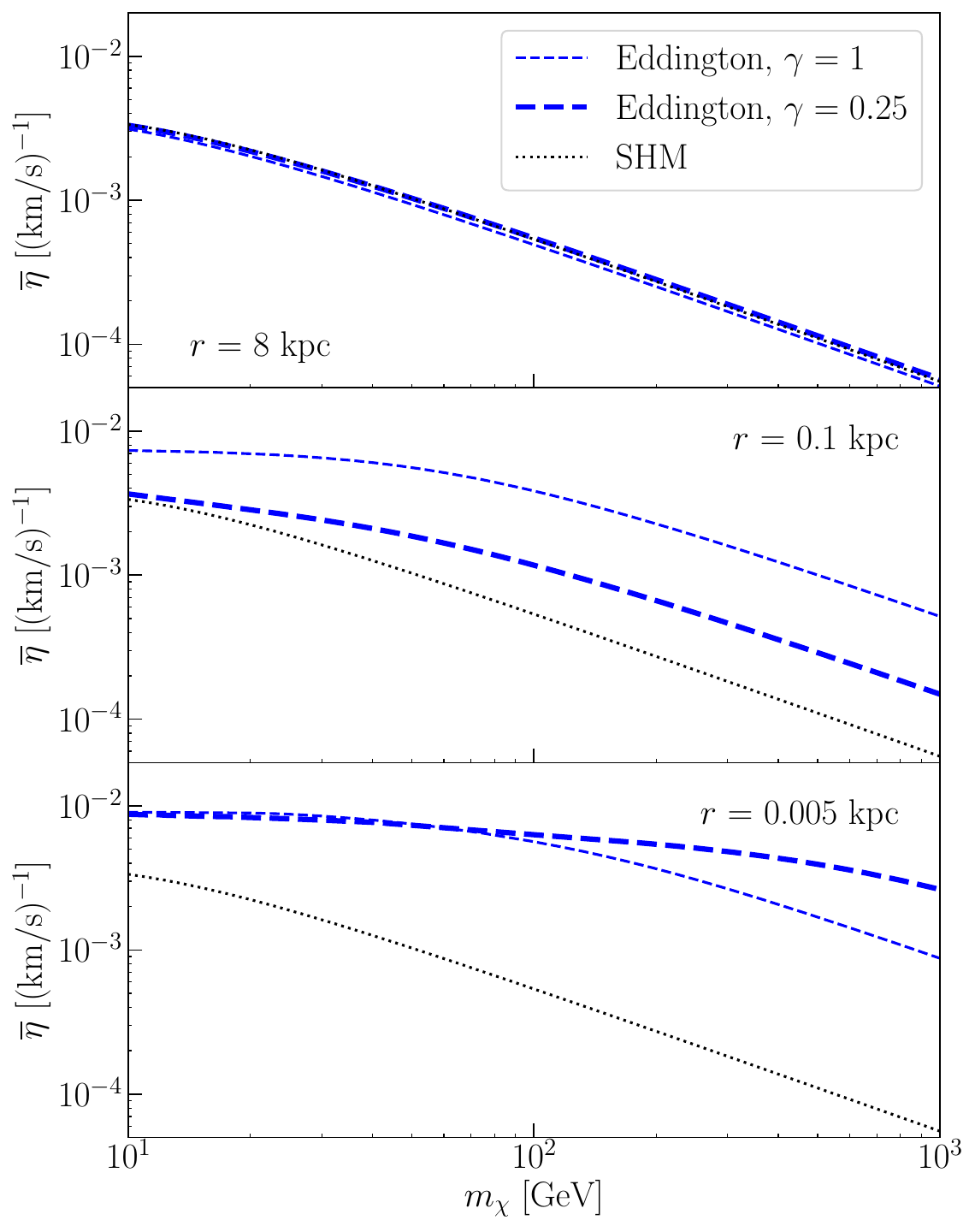}
   \caption{\textit{Left panel:} First inverse moment of the DM velocity distribution $\bar{\eta}$ as a function of $u_{\mathrm{max}}$, at three values of the distance to the center of the Galaxy: 8 kpc (upper panels), 100 pc (middle panels) and 5 pc (lower panels), assuming the SHM (black solid) and Eddington inversion with $\gamma=0.25$ (thin blue dashed) and $\gamma=1$ (thick blue dashed). \textit{Right panel:} Same but as a function of the mass of the DM particle, assuming interactions with hydrogen at the center of the Sun (see Eq.~\ref{eq:u_max}).}
    \label{fig:eta}
\end{figure}

\change{In Sec.~\ref{ssec:capture_formalism} we showed that the first inverse moment of the VDF is the main DM velocity contribution to the capture rate integral (in fact, it is the only contribution when $v_{\mathrm{esc}}\gg u$), which in turn is evaluated between} $u=0$ and $u_{\mathrm{max}}$---the maximum velocity for capture in one collision. Moreover, we also discussed the fact that this maximum velocity, which defines the number of DM particles available for capture, is a function of the DM and target nuclei masses, and decreases as the difference between these two masses increases. In this section we study how the first inverse moment of the speed distribution behaves as a function of the maximum velocity $u_{\mathrm{max}}$. To do this, we use the quantity \cite{Lacroix2018}
\begin{align}
    \bar{\eta}(u_{\mathrm{max}},r) &\equiv \int_0^{u_{\mathrm{max}}} \! \mathrm{d}u \, \frac{f_u^*(r,u)}{u}\,,
    \label{eq:eta}
\end{align}
which essentially reflects---in a somewhat more straightforward way than the full expression in Eq.~\eqref{eq:cap_diff}---the dependence of the capture rate on the DM speed DF. Using the results obtained in Sec.~\ref{sssec:variation_vdf}, we computed $\bar{\eta}(u_{\mathrm{max}},r)$ as a function of the maximum velocity $u_{\mathrm{max}}$ (Fig.~\ref{fig:eta}, left panel) and as a function of the DM particle mass $m_{\chi}$ (Fig.~\ref{fig:eta}, right panel), for the specific case of collisions with hydrogen at the center of the Sun, at three benchmark galactocentric radii: 8 kpc (upper row), 100 pc (middle row) and 5 pc (lower row). As expected, $\bar{\eta}(u_{\mathrm{max}},r)$ increases with the maximum velocity $u_{\mathrm{max}}$, whereas the difference between the SHM and the Eddington model reaches a minimum as $u_{\mathrm{max}} \rightarrow \infty$, \ie, when the integral is performed over the full speed DF domain regardless of the assumed speed DF. More importantly, this difference is more pronounced for $r= 5 \ \mathrm{pc}$ and $r= 100 \ \mathrm{pc}$, being almost negligible at $r= 8 \ \mathrm{kpc}$. This is a direct consequence of the fact that the SHM is no longer valid as we move away from the vicinity of the Sun and consider regions closer to the center of the MW. \change{It should be noted that we did not include the other MB models in Fig.~\ref{fig:eta} since the point of this figure is mainly to discuss the differences between a prototypical MB model and the standard Eddington model, which are representative of these two classes of models for the features discussed in this section.}

In the right panel of Fig.~\ref{fig:eta} we show $\bar{\eta}(u_{\mathrm{max}},r)$ as a function of the mass of the incoming DM particle $m_{\chi}$, assuming only scatterings with hydrogen at the center of the Sun. In this case, which is still representative of the case of scatterings with elements other than hydrogen, we can see that the difference between the SHM and Eddington cases increases with $m_{\chi}$, which means that the impact of phase-space modeling on the capture rate is stronger for heavier incoming DM particles, as we will see in the next section.

Finally, Fig.~\ref{fig:eta} also illustrates how the underlying DM profile, especially the inner slope $\gamma$, plays an important role in shaping the PSDF and subsequently the capture rate, as illustrated by the difference between the thick and thin blue dashed curves in both panels, which grows as one moves towards the GC.

\section{Systematic errors on capture rates from phase-space modeling of DM in the Galaxy}
\label{sec:results}

In this section we revisit predictions for the DM capture rate in different astrophysical scenarios that are relevant in the context of past, current and future indirect DM searches. We consider the PSDFs from both the Eddington method and its anisotropic extension described in Sec.~\ref{sec:first_principles}, and discuss the associated systematic uncertainties on DM constraints from the Sun and NSs, which both depend on the capture rate.

\subsection{Capture in the Sun and neutrinos from DM annihilation}
\label{ssec:capture_sun_neutrinos}

\begin{figure}[t!]
    \centering
    \includegraphics[width=1.0\columnwidth]{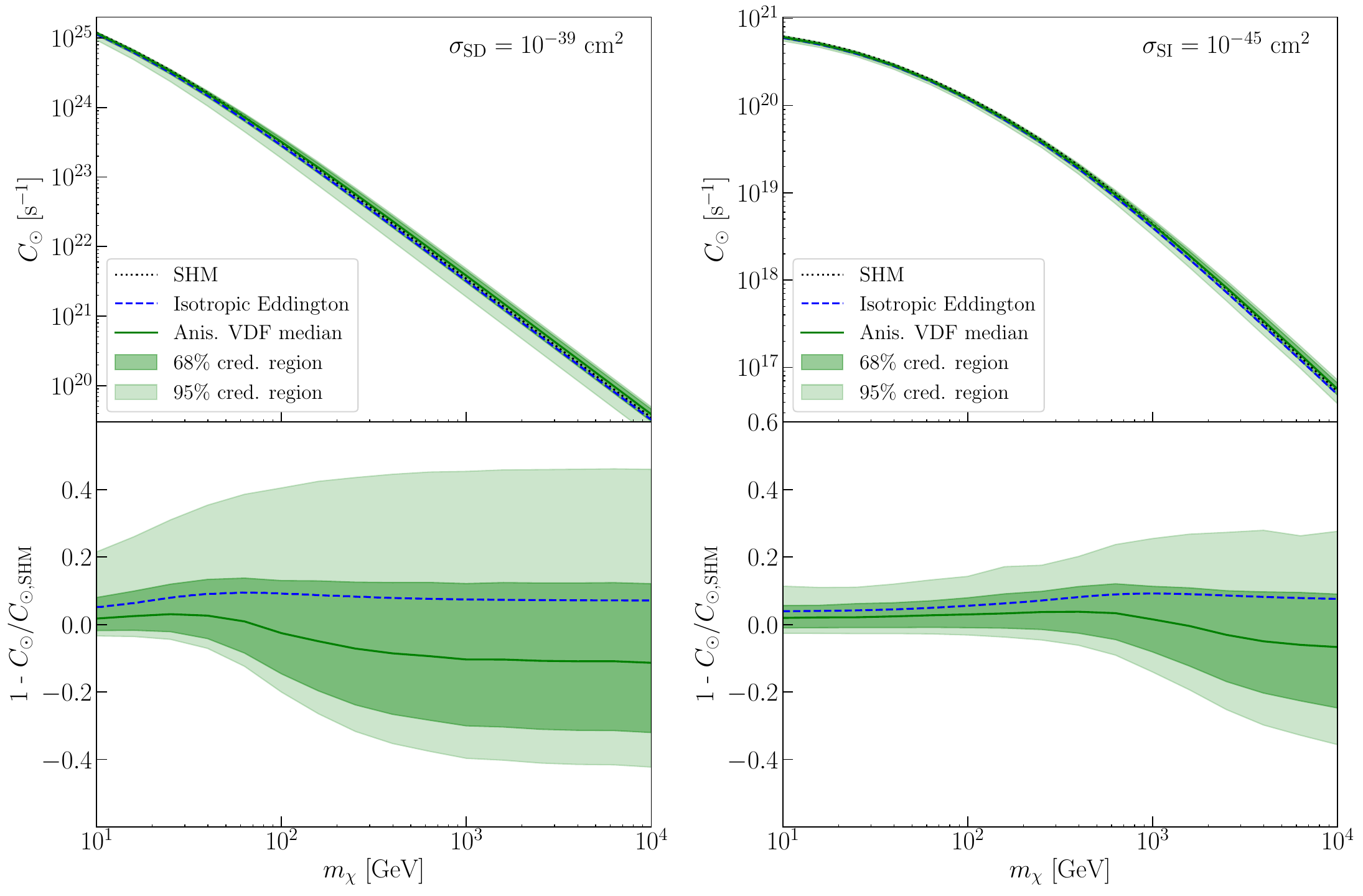}
    \caption{\textit{Top:} Capture rate in the Sun for an NFW DM density profile as a function of DM mass for different models for the speed DF, for SD scattering (left panel), and SI scattering (right panel). The Eddington inversion predictions are shown as dot-dashed blue lines. Solid lines correspond to the median of the anisotropic speed DFs. The green shaded bands represent the associated 68\% and 95\% containment regions, and quantify the uncertainty from the unknown anisotropy. The SHM speed DF, with $v_{\odot} = 220\ \rm km\, s^{-1}$, is shown as dotted lines. \textit{Bottom:} Relative difference between predictions from self-consistent models and the SHM.}
    \label{fig:capture_sun}
\end{figure}

In order to quantify the effect of DM phase-space modeling on constraints from the non-observation of DM-induced neutrinos from the Sun, we computed the capture rate in the Sun as a function of the DM mass $m_{\chi}$ for different speed DFs: (i) the SHM, which is used in most studies in the literature, (ii) the speed DF from the Eddington inversion, and (iii) the anisotropic speed DFs obtained by marginalizing over the anisotropy parameters, as described in Sec.~\ref{ssec:uncertainty_anisotropy}.\footnote{\change{At the location of the Sun the MB models with $v_0 = v_{\rm c}(r)$ and $v_0 = \sqrt{2/3} \sigma_{v}^{\rm J}(r)$ are almost identical to the SHM so the results we obtain from capture in the Sun are representative of all MB models discussed in this work.}} The results are shown for the NFW profile, as a function of the DM mass in Fig.~\ref{fig:capture_sun} for SD (left panel) and SI (right panel) scattering, along with the relative difference with respect to the standard values of the capture rate obtained with the SHM in the lower panels. The Eddington and SHM predictions are shown as blue dot-dashed and black dashed lines, respectively. In the anisotropic case, the capture rate was computed for values of the $(\beta_0,\beta_\infty, L_0)$ parameters allowed by dynamical constraints (see Sec.~\ref{ssec:uncertainty_anisotropy}). The result marginalized over these parameters is given by the median (green solid line) and the 68\% and the 95\% credibility regions (green shaded bands) which quantify the uncertainty on the capture rate from the unknown anisotropy. We used a solar model obtained with the MESA code \cite{Paxton2011, Paxton2013, Paxton2015, Paxton2018, Paxton2019}, and assumed a local DM density of $\rho_{\mathrm{DM}} = 0.38\, \rm GeV\, cm^{-3}$, which is not only consistent with the McM17 mass model, but also with recent local DM density estimates (\eg
~\cite{Catena2010}).\footnote{\change{This value is also essentially consistent with the slightly larger values found in refs.~\cite{Salucci2010,Nesti2013}.}} For the interaction cross-sections we adopted the benchmark values for the cross sections $\sigma_{\mathrm{SD}} = 10^{-39}\, \mathrm{cm}^2$ and $\sigma_{\mathrm{SI}} = 10^{-45}\, \mathrm{cm}^2$, which are representative of current experimental limits (\eg~\cite{pdg2018}). All speed DFs are boosted to the solar frame, assuming a circular velocity of $v_\odot \approx 220\, \mathrm{km\, s^{-1}}$.\footnote{\change{$v_\odot = v_{\rm c}(8\ \rm kpc) = 221\, \mathrm{km\, s^{-1}}$ for the sphericized NFW McM17 mass model.}} For the SHM we used the usual value of the most probable speed $v_0 = v_\odot = 220\, \mathrm{km\, s^{-1}}$.

\begin{figure}[t!]
    \centering
    \includegraphics[width=1.0\columnwidth]{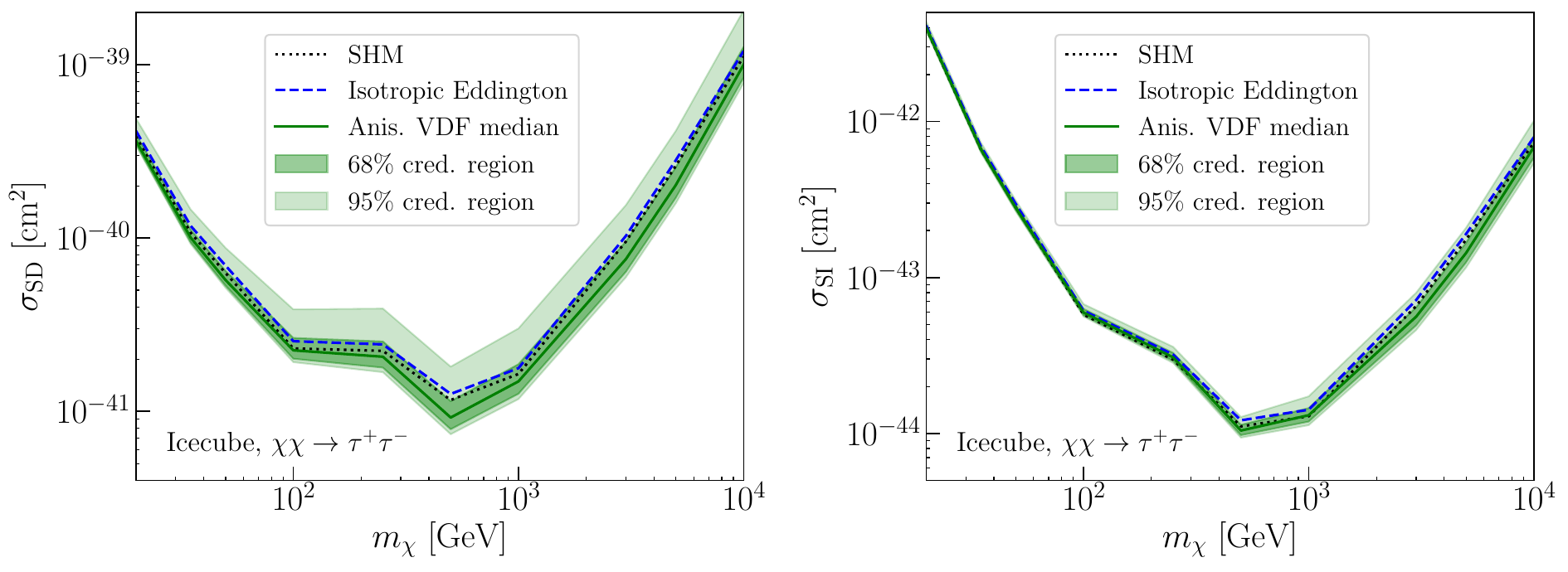}
    \caption{Upper limits on the DM SD (left panel) and SI (right panel) WIMP-nucleon scattering cross section obtained in this work from null results from IceCube \citep{aartsen2017} for each different model for the speed DF. Line styles are the same as in Fig.~\ref{fig:capture_sun}. See text for details.}
    \label{fig:nu_limits}
\end{figure}

The capture rates computed with the SHM and the Eddington inversion lie within 10\% of each other, which is somewhat expected given the similarity between both speed DFs at $r = 8 \ \mathrm{kpc}$ (see Fig.~\ref{fig:fu_boosted}). This is consistent with the fact that the solar neighborhood corresponds to the galactocentric radius where the DM halo is expected to be closest to isothermal. This relative error turns out to be of the same order as the theoretical error on the Eddington prediction itself quantified in ref.~\cite{Lacroix2020} through self-consistent comparisons with simulations.

The green shaded bands in Fig.~\ref{fig:capture_sun} show that the uncertainty on the anisotropy of the DM velocity tensor leads to an error of $\sim 40\%$ and $\sim 20\%$ (for SD and SI scattering, respectively) on the capture rate over the WIMP mass range. This gives an estimate of the systematic uncertainty on capture in the Sun from DM phase-space modeling, which is not accessible when restricting the predictions to the SHM.

To quantify the systematic uncertainty on actual DM constraints from neutrino telescopes and deviations from the SHM, we computed upper limits on the DM-nucleon scattering cross section, using the upper limits on neutrino fluxes from the Sun from the IceCube Collaboration \cite{aartsen2017}. The results are shown in Fig.~\ref{fig:nu_limits} for SD and SI interactions in the left and right panels, respectively. Only the limits for the $\chi \chi \rightarrow \tau^{+} \tau^{-}$ are shown since they are the most stringent and representative of the other annihilation channels. As in other neutrino studies, we assume that the number of DM particles in the Sun has reached a steady state \cite{jungman1996}. Then, as long as the cross section is much smaller than the geometric limit (which is the case here), $C_{\odot}$ scales linearly with the cross section (see Eq.~\ref{eq:cap_diff}), and the relative difference of the neutrino limits for the different speed DFs reflects the relative differences on the capture rate illustrated in Fig.~\ref{fig:capture_sun}. In particular, the limits for the SHM and Eddington inversion have a maximum difference of $\sim 10\%$ for both SD and SI interactions and are both contained within the 68\% credibility region of the allowed anisotropy configurations. Although the SHM indeed provides a reasonable approximation of the galactic speed DF in the vicinity of the Sun, \ie, at $r \simeq 8 \ \mathrm{kpc}$, the reconstruction of the anisotropic PSDF allows us to estimate a realistic systematic uncertainty on the neutrino limits of a factor 2 for SD scattering (of order 40\% for SI scattering) from phase-space modeling.\footnote{It should be noted that the median of the capture rate in Fig.~\ref{fig:capture_sun} for the anisotropic model is the result of the marginalization over the anisotropy parameters and does not correspond to a specific speed DF. Therefore, we do not use directly the corresponding values of the capture rate to derive limits on the DM-nucleon cross section from the non-observation of neutrinos from the direction of the Sun with IceCube. Instead, we compute the limits for all the anisotropic models and marginalize afterwards, which leads to the 68\% and 95\% credible regions in Fig.~\ref{fig:nu_limits}.} This is consistent with the results of ref.~\cite{Ibarra2018}, where the authors introduced an additional parameter to account for departures from the SHM in the Solar neighborhood, and used it to derive the maximum allowed variations of neutrino and direct detection limits. We emphasize that our results on capture in the Sun are based on PSDFs obtained with the anisotropic extension of the Eddington formalism. We do include the result for the isotropic Eddington model in the plots for reference---although it was also presented in ref.~\cite{Nunez2019}---to illustrate how the anisotropic model allows us to go a step further when evaluating uncertainties on DM-capture related observables.

Finally, it should also be noted that there is a small decrease in the relative difference on the capture rate in Fig.~\ref{fig:capture_sun} (and equivalently on the upper limits in Fig.~\ref{fig:nu_limits}) for small $m_{\chi}$. This is due to the effect already described in Sec.~\ref{sssec:maxvel_dep}: by decreasing the mass difference between the incoming DM particle and the target nucleus, we are effectively probing all the population of DM particles available in the Galactic halo, which smooths out differences in the speed DF. This also explains the difference between the SI and SD cases, since, in comparison with SI interactions, SD scatterings with elements heavier than hydrogen are highly suppressed. Therefore, for a given DM mass, in SI capture---which in the Sun is dominated by $^{4}\mathrm{He}$ and $^{16}\mathrm{O}$---the maximum velocity for capture $u_{\rm max}$ (see Eq.~\ref{eq:u_max}) is larger than for SD interactions. As a result, the relative difference between capture rate predictions in the SI case has essentially the same behavior as in the SD case, but shifted to higher masses.

To conclude this section, we have shown that Eddington-like methods make it possible to quantify systematic uncertainties associated with the underlying properties of the PSDF of DM on neutrino limits from capture in the Sun. Although the SHM \change{and MB models in general provide} a good estimate of the constraints in the solar neighborhood, the actual limits may differ in a non-negligible way from the results \change{that rely on the MB distribution}, which is crucial for instance when translating such limits into constraints on DM models.

\subsection{Capture rates for solar-like stars at different positions in the Galaxy}
\label{ssec:capture_sun_like}

Careful modeling of the PSDF of Galactic DM becomes increasingly important as we consider observables that probe DM closer to the GC, where the DM halo is expected to strongly depart from the isothermal regime associated with the SHM. It is important to note that due to current experimental limitations, it is extremely hard to probe the DM content of stars in the central regions of the Galaxy. Nonetheless, several studies have explored different strategies to use stars in these regions to search for convincing DM signals \citep{scott2009,lopes2019,lopes2019b}. Furthermore, projected experimental efforts aimed at the GC should eventually unlock the full potential of stars in this region as testing grounds for DM search, given the high DM densities expected there.

Anticipating such promising opportunities, in this section we study the dependence of the DM capture on the underlying speed DF at different locations in the MW. We consider a star identical to the Sun as if it were located in different regions of the Galaxy. The reason why we consider a solar-like star (in this case, an exact replica of the Sun), and not for instance any heavier stars, is because most DM effects and signatures are negligible in the latter, due to the explosiveness and instability that characterizes their evolution \cite{scott2009}. It should also be noted that in this section we focus on regular stars for which the phenomenology of DM capture is expected to be different from that of NSs, which we discuss in Sec.~\ref{ssec:ns}.

\begin{figure}
    \centering
    \includegraphics[width=1.0\columnwidth]{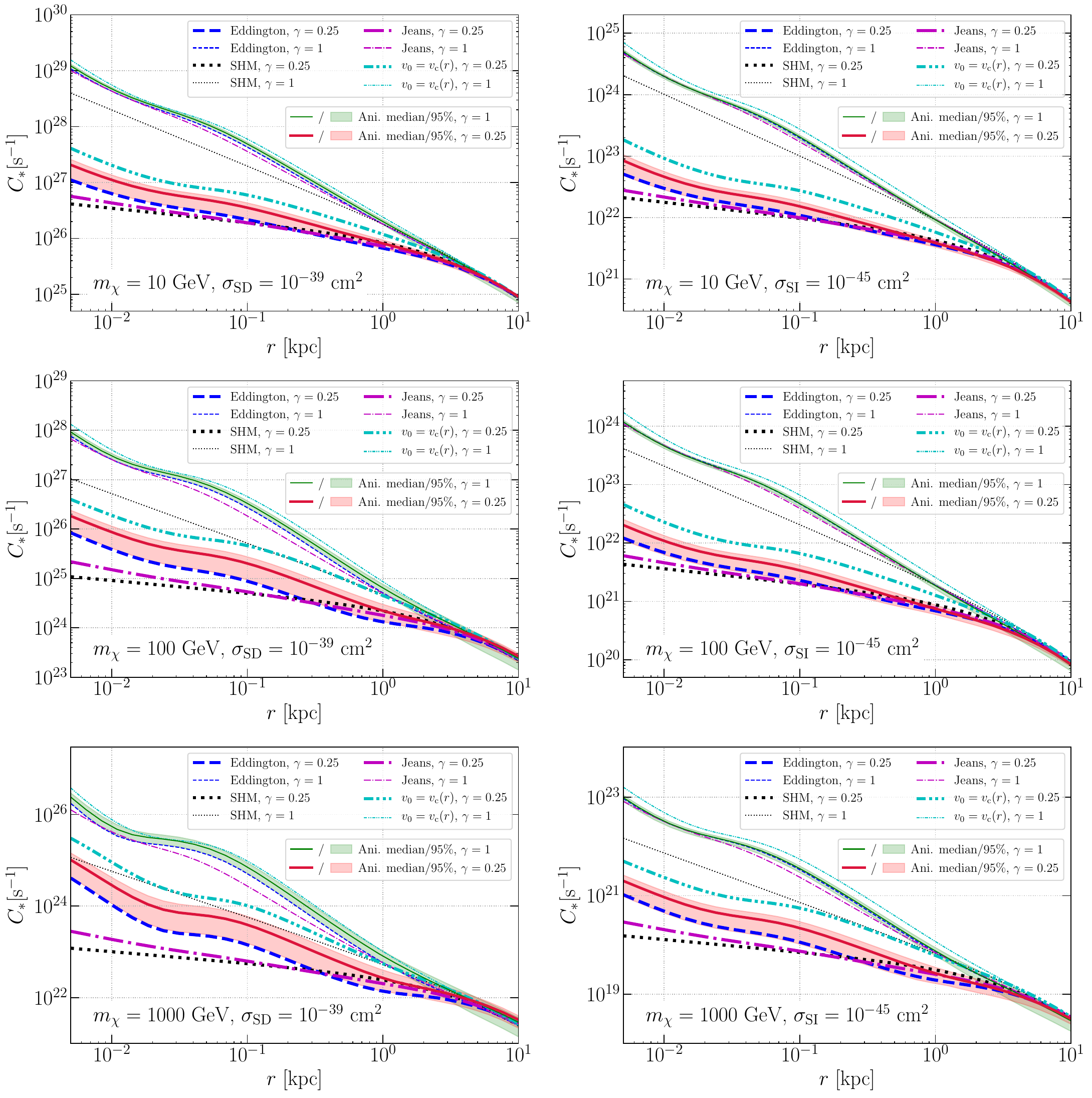}
    \caption{Capture rate in a solar-like star as a function of galactocentric radius in the MW for DM masses of 10~GeV (upper panels), 100~GeV (middle panels) and 1~TeV (bottom panels), for SD (left) and SI (right) interactions. \change{Predictions from the Eddington inversion are shown as blue dashed lines. For the anisotropic models the median is displayed as a solid line and 95\% confidence region as a shaded band, green for $\gamma = 1$ and red for $\gamma = 0.25$. Predictions of MB models with $v_0 = v_{\odot} = 220\ \rm km\, s^{-1}$ (\ie~the SHM), $v_0 = v_{\rm c}(r)$, and $v_{0} = \sqrt{2/3} \sigma_{v}^{\rm J}(r)$ are shown as black dotted, cyan dot-dot-dashed, and magenta dot-dashed lines, respectively. Thin lines correspond to a cored DM profile with $\gamma = 0.25$, and thick lines to a cuspy profile with $\gamma = 1$.} Capture rates are computed for benchmark values of $10^{-39}\, \rm cm^{2}$ and $10^{-45}\, \rm cm^{2}$ for the DM-nucleon cross section for SD and SI interactions, respectively.}
    \label{fig:capture_rate_profile_sun_like}
\end{figure}

We computed the capture rate for a solar-like star as a function of galactocentric radius, for the \change{PSDFs presented in Sec.~\ref{sssec:variation_vdf}, namely from the Eddington inversion method and its anisotropic extension, and MB distributions with $v_0 = v_{\odot} = 220\ \rm km\, s^{-1}$ (\ie~the SHM), $v_0 = v_{\rm c}(r)$, and $v_{0} = \sqrt{2/3} \sigma_{v}^{\rm J}(r)$.} As discussed in Sec.~\ref{ssec:mass_models}, we considered both the usual NFW density profile with inner slope $\gamma = 1$ and a cored profile with $\gamma = 0.25$ in order to bracket the uncertainty on the inner slope. The results are shown in Fig.~\ref{fig:capture_rate_profile_sun_like}, for DM particle masses of 10 GeV, 100 GeV and 1 TeV (upper, middle, and lower panels, respectively) for SD and SI interactions (left and right panels, respectively). Clearly the largest systematic uncertainty comes from the poorly constrained inner slope $\gamma$ of the DM density profile, with differences of up to 2 orders of magnitude between the results for cored and cuspy profiles. It should be noted that the dependence of the capture rate on $r$ is not only due to the density profile $\rho_{\chi}(r)$---the capture rate in all scenarios increases when moving closer to the GC, more strongly so for $\gamma = 1$ than for $\gamma = 0.25$---but also to variations of the speed DF. While for the SHM with fixed solar neighborhood values the capture rate is directly proportional to the DM density (see Eq.~\ref{eq:cap_diff})\change{, the other MB models, as well as} Eddington-like models are able to capture features in phase space caused by the various components of the MW. The most prominent here are the bulge at $r = \mathcal{O}(100\, \mathrm{pc})$ and the nuclear star cluster (or inner bulge) at $r = \mathcal{O}(5\, \mathrm{pc})$ (see Table~\ref{mass_model_table} in App.~\ref{appendixB}), which lead to characteristic speeds at a certain scale.

As shown in Fig.~\ref{fig:capture_rate_profile_sun_like}, the SHM increasingly underestimates the capture rate with decreasing $r$. This can be easily understood by looking at the underlying VDFs shown in Fig.~\ref{fig:fu_boosted}: whereas (by construction) the most probable velocity in the SHM is the same for every $r$, for the Eddington model and its anisotropic extension the boosted speed DFs peak at lower speeds, therefore enhancing the capture rate (incoming particles with lower speed are indeed easier to capture). Moreover, this difference is larger for heavier particles, which, again, is a consequence of the relationship between the maximum speed for a particle to be captured in a single collision, and the mass difference between the incoming particle and the target nucleus (see Sec.~\ref{sssec:maxvel_dep}). Fig.~\ref{fig:capture_rate_profile_sun_like} shows that, as we approach the center of the MW, the SHM underestimates the capture rate in comparison with models that self-consistently account for the dynamics of the Galaxy. These differences can get significant for larger masses, reaching one order of magnitude for $m_{\chi} = 1\, \mathrm{TeV}$ at $r \lesssim 10\, \mathrm{pc}$ for $\gamma = 1$ and up to two orders of magnitude for a cored profile with $\gamma = 0.25$. Moreover, for the same $r$ and mass values, the uncertainty on the capture from the unknown underlying anisotropy of the speed DF amounts to $\sim 75\%$ for $\gamma = 1$ and up to a factor of $4$ for $\gamma = 0.25$. It should be noted that the difference in the anisotropy bands between different inner slopes $\gamma$ for small $r$ is related to the underlying density profile and is consistent with what was obtained in Sec.~\ref{sssec:variation_vdf}. The same goes for the asymmetry between the isotropic case and the anisotropy bands in the same region.

\change{Regarding the alternative MB models discussed in Sec.~\ref{sssec:variation_vdf}, when considering $v_0 = v_{\rm c}(r)$ one overestimates the low-velocity tail of the PSDF at small radii (even more so for the $\gamma = 1$ case) with respect to Eddington-like models, and thus overestimates the capture rate by a factor $\sim 2$ for $\gamma = 1$, and up to a factor $\sim 4$ for $\gamma = 0.25$, as illustrated by the cyan dot-dot-dashed lines in Fig.~\ref{fig:capture_rate_profile_sun_like}. When using the MB model with velocity dispersion obtained by solving the Jeans equation, the capture rate lies much closer to the Eddington result than the other MB models for the $\gamma = 1$ profile, but it is very close to the SHM result for the cored profile with $\gamma = 0.25$, as shown by the magenta dot-dashed lines in Fig.~\ref{fig:capture_rate_profile_sun_like}.}

These results show that by using the SHM to predict the capture rate for stars in the inner regions of the Galaxy one systematically underestimates any potential capture-related DM signature. \change{On the other hand, the capture rate is systematically overestimated when using a MB model peaking at the circular speed $v_{\rm c}(r)$. We also showed that the MB model with the Jeans velocity dispersion appears as a good proxy for the more physically motivated Eddington model for a cuspy profile (although it still underestimates the capture rate), but for a cored profile it does not represent a significant improvement over the SHM. This reflects the intrinsic limitations of MB models which are by construction unable to account for non-Maxwellian features in the PSDF introduced in particular by the baryonic gravitational potential. On the contrary, the anisotropic Eddington-like model allows us to simultaneously account for departures from a Gaussian distribution, as well as the characteristic speeds dictated by the dynamics of the MW at every radius, and the uncertainty from the unknown anisotropy of the DM velocity tensor, while retaining only models that provide stable solutions to the collisionless Boltzmann equation---as discussed in Sec.~\ref{ssec:uncertainty_anisotropy}.}

\subsection{Neutron stars}
\label{ssec:ns}

Due to the presence of a nuclear star cluster at the GC, a large number of NSs is expected to exist in that region. \change{While the observation of a DM signature in the NS cooling process is technologically restricted to NSs in the solar neighborhood, constraints from gravitational stability considerations are expected to be more stringent in regions of high DM density, which is presumably the case for the GC (although this depends on the actual underlying DM density profile)}. These two factors make the GC a prime target to study signatures of DM capture in NS. As yet, NSs from the very inner regions of the Galaxy have not been observed, but many projected constraints do rely on studying NSs in regions of DM densities of at least $10^{3}\, \rm GeV\, cm^{-3}$ (\eg~\cite{Garani2019}). However, in these regions the DM PSDF is expected to depart much more significantly from the SHM\change{---and more generally from a MB distribution---}than in the Solar neighborhood, so being able to robustly estimate PSDF-related uncertainties is even more crucial.

Nevertheless, while on the one hand a tremendous amount of work has been done to reduce theoretical uncertainties on capture in NSs by accounting for several effects that were not included in seminal papers on the derivation of capture rates---such as corrections from multiple scatterings from large baryonic densities in NSs which can boost overall capture rates especially for DM particles with $m_{\chi} \gtrsim 100\, \mathrm{TeV}$ (\eg~\cite{Dasgupta2019}), or effects of the degeneracy of the NS medium which can limit the capture rate due to the Pauli exclusion principle \cite{Garani2019}---on the other hand models of the Galactic PSDF of the DM particles have always remained very simple and the associated systematic errors have not been discussed. In particular, in refs.~\cite{Kouvaris2008,Kouvaris2010,McDermott2012,Guver2014,Bramante2013,Bertoni2013,Bramante2014,Bramante2017,Garani2019}, the authors relied on a MB model with most probable speed fixed at $v_{0} = 220\, \rm km\, s^{-1}$ everywhere in the galaxy (\ie~the SHM), even when making projections for NSs in regions of much higher DM density than the Solar neighborhood. In refs.~\cite{Bertone2008,Bell2013,BramanteLinden2014,Bramante2018,Gresham2019,Dasgupta2019}, the discussion goes a bit further, leaving room for a rescaling of the results as a function of the typical DM speed. However, this still ignores the important correlation between the DM density and the speed DF encapsulated in the full PSDF. An exception is ref.~\cite{deLavallaz2010}, where the authors \change{did solve the Jeans equation and considered a MB model with $v_0 = \sqrt{2/3}\, \sigma_{v}^{\rm J}(r)$. This goes beyond the usual approach relying on the SHM. We note that in that work the authors} focused on (sub)-parsec scales so the central black hole played an important role, which is somewhat more specific than the general picture we are trying to provide here.

It is important to note that, as previously discussed in the literature, for a typical NS, the maximum speed for capture in a single collision is $u_{\rm max} \sim 10^{5}\, \rm km\, s^{-1}$, which is much larger than the typical speed of DM particles, of a few $10^{2}\, \rm km\, s^{-1}$. As a result, this has been often used in the literature as an argument to claim that the DM speed DF should not have any impact on the capture rate (see also Sec.~\ref{ssec:vel_space_capture}). Still, in this work we quantify this effect for the first time and show in the following that it is definitely non-negligible even for NSs. Therefore, even if in some DM mass ranges phase-space effects are bound to be much smaller than other effects, such as suppression of the capture rate from the Pauli exclusion principle, they need to be accounted for in capture analyses.

\begin{figure}[t!]
    \centering
    \includegraphics[width=0.75\columnwidth]{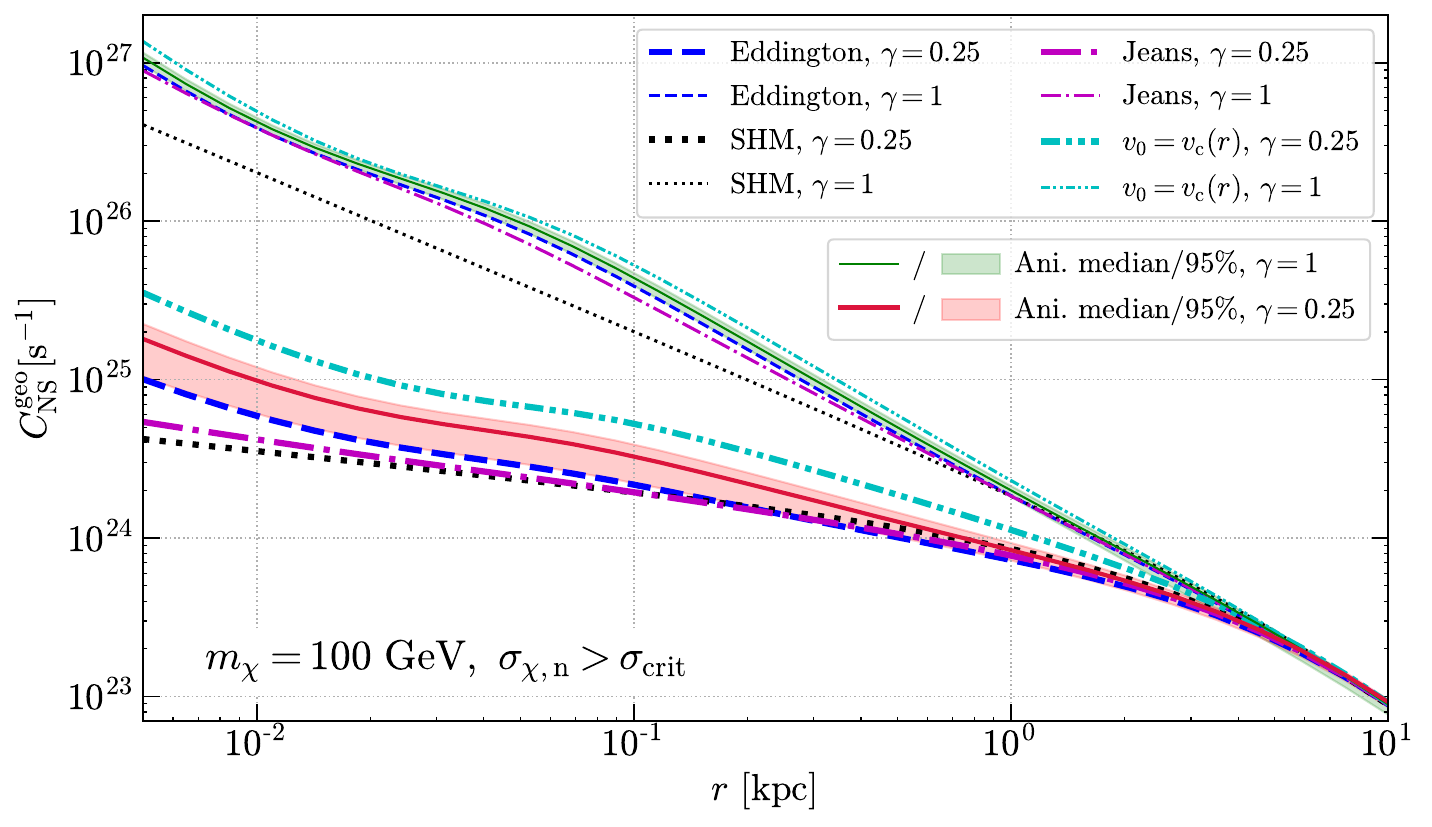}
    \caption{Capture rate in the geometric limit for a NS with $M_* = 1\, \rm M_{\odot}$ and $R_* = 10\, \rm km$ as a function of the distance to the center of the MW. The various line styles are the same as in Fig.~\ref{fig:capture_rate_profile_sun_like}.}
    \label{fig:cap_ns}
\end{figure}

As mentioned in Sec.~\ref{ssec:capture_formalism}, there is an upper limit to the capture rate\change{, namely the geometric limit}, corresponding to the scenario where all DM particles crossing the star are captured (see Eq.~\ref{eq:cap_geo}). \change{When this is the case, the capture rate} is independent of \change{the kinematics of the scattering process, however it is still subject to GR corrections as discussed in Sec.~\ref{ssec:capture_formalism}.}\footnote{As mentioned before, there are other effects due to the degeneracy of the NS medium which can limit the capture rate due to the Pauli exclusion principle. These corrections are negligible for $m_{\chi} \gtrsim 1\, \mathrm{GeV}$ \cite{Garani2019}, and therefore we disregard them in this work.} For a typical NS with $M_* = 1\, \rm M_{\odot}$ and $R_* = 10\, \mathrm{km}$, this situation occurs when $\sigma_{\chi,\mathrm{n}} > \sigma_{\mathrm{crit}} \approx 2.6 \times 10^{-45}\, \mathrm{cm}^2$, which is in line with the cross section considered in this work thus far. For this reason, and given that the scope of this work is to study the effects of phase-space modeling on capture in stars, we only consider the capture rate in the geometric limit. We should note however that the effects observed in the geometric capture due to the speed DF should be representative of the more general case when $\sigma_{\chi,\mathrm{n}} < \sigma_{\mathrm{crit}}$.

\change{In Fig.~\ref{fig:cap_ns} we show the geometric capture rate in a typical NS computed assuming MB distributions with $v_0 = v_{\odot} = 220\ \rm km\, s^{-1}$ (\ie~the SHM), $v_0 = v_{\rm c}(r)$, and $v_{0} = \sqrt{2/3} \sigma_{v}^{\rm J}(r)$, as well as both the isotropic and anisotropic Eddington inversion speed DFs. The absolute values of the capture rates shown in Fig.~\ref{fig:cap_ns} should not be taken at face value since they do not account for GR corrections, but we provide them for the sake of definiteness and reproducibility, and they still allow us to illustrate the variation of the capture rate depending on the phase-space model. In the following, when discussing actual limits, we consider ratios between different phase-space models, for which GR factors cancel out.} We only show the results for $m_{\chi} = 100\, \mathrm{GeV}$ since, in the mass range relevant for this work ($10\, \mathrm{GeV} \lesssim m_{\chi} \lesssim 10^5\, \mathrm{GeV})$, $C_{\mathrm{NS}}^{\mathrm{geo}}$ scales linearly with $1/m_{\chi}$. The relative behavior between the capture rate obtained with the \change{MB models} and with Eddington-like speed DFs is analogous to the solar-like case studied in the last section. The reason why the impact of the Eddington inversions PSDF on the capture rate is typically smaller for NS than for what solar-like stars (see Fig.~\ref{fig:capture_rate_profile_sun_like}), is that in a NS the maximum velocity for capture is much larger than the typical velocities of DM particles in the Galactic halo as discussed above. Because of this, the integral in Eq.~\eqref{eq:cap_geo} spans the entire velocity range available, \change{mitigating} any region-specific differences between speed DFs. This is valid for all relevant values of $m_{\chi}$ due to the compact nature of NS.

\begin{figure}[t!]
    \centering
    \includegraphics[width=1.0\columnwidth]{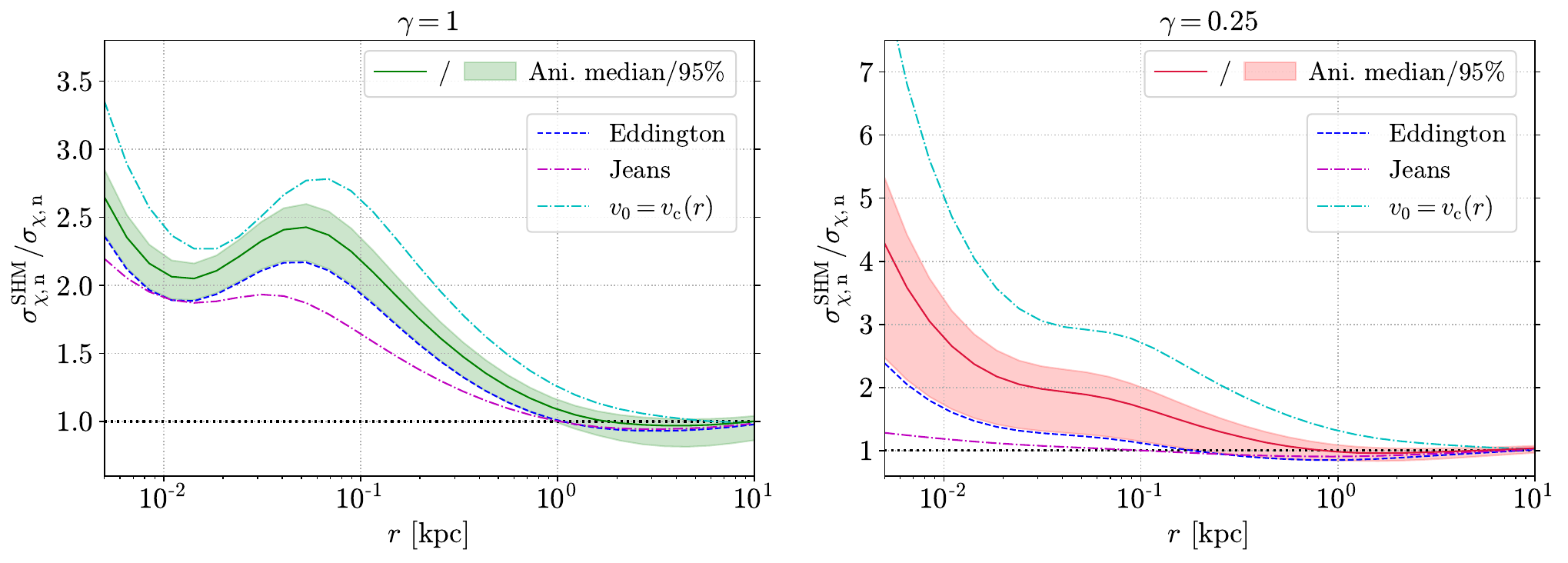}
    \caption{\change{Ratios of upper limits on the DM-neutron scattering cross section from core collapse criteria obtained using the SHM to those obtained with more realistic speed DFs, for stars at different radii $r$. Results are shown for a NFW density profile corresponding to $\gamma = 1$ (left) and a cored profile with $\gamma = 0.25$ (right). Predictions from the Eddington inversion are shown as a blue dashed line. For the anisotropic models the median is displayed as a solid line and the 95\% confidence regions as a shaded band, green for $\gamma = 1$ and red for $\gamma = 0.25$. Predictions of MB models with $v_0 = v_{\rm c}(r)$ and $v_{0} = \sqrt{2/3} \sigma_{v}^{\rm J}(r)$ are shown as before as cyan dot-dot-dashed and magenta dot-dashed lines, respectively. The limits for NSs close to the GC are projections.}}
    \label{fig:ns_lim}
\end{figure}


\change{Following the reasoning presented in ref.~\cite{Garani2019}, upper limits on the DM-neutron cross section can be obtained for a NS located at a given galactocentric radius. The NS capture rate in the non-geometrical limit, \ie~for $\sigma_{\chi,\mathrm{n}} < \sigma_{\mathrm{crit}}$, can be obtained by rescaling $C_{\mathrm{NS}}^{\mathrm{geo}}$, which is a valid approach since in the mass range of interest, $1-10^5 \ \mathrm{GeV}$, the capture rate scales linearly with $\sigma_{\chi,\mathrm{n}}$. More specifically, this translates to \begin{equation}
    C_{\rm NS} = C_{\mathrm{NS}}^{\mathrm{geo}} \dfrac{\sigma_{\chi,\mathrm{n}}}{\sigma_{\mathrm{crit}}}\,.
\end{equation}
Ratios of limits computed for different PSDF models, but for the same physical conditions in the NS, are then given by the corresponding ratios of the geometric capture rates. So here we consider ratios of the limits for the SHM to limits for the other PSDF models considered in this work:
\begin{equation}
    \dfrac{\sigma_{\chi,\mathrm{n}}^{\rm SHM}}{\sigma_{\chi,\mathrm{n}}} = \dfrac{C_{\mathrm{NS}}^{\mathrm{geo}}}{C_{\mathrm{NS}}^{\mathrm{geo,SHM}}}\,.
\end{equation}
The results are shown in Fig.~\ref{fig:ns_lim}, for $\gamma = 1$ (left panel) and $\gamma = 0.25$ (right panel), for the Eddington PSDF and its anisotropic extension, and the MB models with $v_0 = v_{\rm c}(r)$ and $v_0 = \sqrt{2/3} \sigma_{v}^{\rm J}(r)$.} While for $r = 8\ \mathrm{kpc}$ there is no obvious difference between the limits obtained with the Eddington-like speed DFs and the SHM, as we consider stars closer to the center of the MW, the SHM systematically underestimates the ability of NS to constrain DM-neutron scattering, which is consistent with the results obtained in Fig.~\ref{fig:cap_ns}. For instance, for a NS at $r = 5 \, \mathrm{pc}$, compared with the Eddington-like models, the SHM overestimates the upper limits on $\sigma_{\chi,\mathrm{n}}$ by a factor $\sim 2.5$ for $\gamma = 1$, and 2.5 (4 when considering the median of the anisotropic models, solid line) for $\gamma = 0.25$. \change{The situation is slightly different when considering the MB-Jeans model with $v_0 = \sqrt{2/3} \sigma_{v}^{\rm J}(r)$, which follows the Eddington result within a few tens of \% for $\gamma = 1$. However, for $\gamma = 0.25$ the MB-Jeans model remains very close to the SHM, but underestimates the Eddington result slightly less than the SHM, with a difference smaller than a factor 2. As for the $v_0 = v_{\rm c}(r)$ model, it systematically overestimates the limits by a factor 2-3, so it does not represent an improvement over the SHM, and can even bias the limits even further in the case of a cored profile.}

\change{In addition to systematic differences between MB and Eddington-like models, uncertainties due to anisotropy account for a variation of $\sigma_{\chi,\mathrm{n}}$ at the level of $\sim 20\%$ for $\gamma = 1$ and $\sim 50\%$ for $\gamma = 0.25$.} 

\change{In this section, we have therefore quantified the robustness of NS constraints with respect to phase-space-related uncertainties. In particular, we have discussed the quantitative differences between results obtained with MB models and full phase-space models from first principles.}




Finally, it should be noted that the results shown in Fig.~\ref{fig:ns_lim} were obtained assuming an equilibrium phase-space model for the DM in the Galaxy---as done systematically in the literature---which is not strictly valid when considering integrated DM capture over long time scales like the age of a NS, which we take here to be 10~Gyr. However, numerical simulations of MW-like systems seem to indicate that about 40\% of the current mass of the DM halo was already present 10~Gyr ago, and about 60\% 8~Gyr ago \cite{Santistevan2020}. Moreover, the recent past of the MW is likely to have been quiet, with evidence for the last major merger about 12~Gyr ago \cite{Wyse2001,Forero-Romero2011}. As a result, over the history of a NS we do not expect a dramatic deviation between the capture rate estimated with a `static' MW model and the true result accounting for the full formation history of the Galaxy---but this would still warrant a dedicated study.\footnote{\change{The \textit{Gaia} results are actually opening a new avenue to investigate Galactic archaeology, as discussed for instance in ref.~\cite{Necib2020}.}} Furthermore, due to the mild dependence of the capture rate in NS on the underlying speed DF which we have quantified here, we can reasonably conclude that the limits on DM-neutron scattering cross sections obtained with equilibrium phase-space models provide a good estimate of the actual limits one would obtain by accounting for the whole history of the Galaxy. This reinforces NSs as very good targets for capture-related DM searches as far as uncertainties from DM phase-space modeling are concerned.

\section{General summary and outlook}
\label{sec:conclusion}

In this work, in an effort to add some quantitative arguments to the discussion of the dependence of dark matter capture on the underlying velocity distribution function, we have estimated systematic uncertainties, from dark matter phase-space modeling, on dark matter constraints based on capture in stars, by using Eddington-like equilibrium phase-space models. These models are based on first principles and their main ingredients are the dark matter density profile and the total gravitational potential, so they self-consistently account for kinematic constraints on the Milky Way. As a result, although they are based on simplifying assumptions, these models are physically motivated and provide a next-to-minimal picture of the properties of the Galactic phase-space relevant to dark matter searches, with respect to standard approaches used in the dark matter literature. These models essentially address several problems at once: first the underlying mass model of the Milky Way constrained by kinematic measurements is by construction accounted for, so that the resulting phase-space distribution function is able to capture variations of the typical dark matter speed as a function of galactocentric radius, in a self-consistent way. \change{Moreover, departures from the Maxwell-Boltzmann approximation, which is at the heart of the standard halo model, are automatically taken into account. Furthermore,} the unknown anisotropy of the dark matter velocity tensor, which is an additional source of uncertainties on the phase-space distribution function, can be accounted for with these self-consistent prediction methods. This goes beyond approaches that empirically quantify departures from the standard halo model in the solar neighborhood, and provides a global physical picture of the dark matter phase space in the Galaxy and the associated systematic uncertainties on dark matter-capture observables. Our main results are the following:

\begin{itemize}
\item We have shown that using a Maxwell-Boltzmann distribution with a very simple estimate of $220\, \rm km\, s^{-1}$ for the typical dark matter speed---the standard halo model---in a region where it is expected to be significantly different based on the underlying mass content---typically at the Galactic center---leads to significant errors on subsequent results. For a Sun-like star within a few pc of the Galactic center, the capture rate can be underestimated by up to almost two orders of magnitude---depending on the dark matter density profile, dark matter candidate mass and the type of interaction with ordinary matter---compared with predictions from Eddington-like models. Even for neutron stars, which are less sensitive to the speed distribution, the capture rate can be underestimated by a factor $\sim 2.5$-4, thus overestimating subsequent upper limits on the dark matter-neutron scattering cross section by the same amount. This definitely needs to be taken into account and cannot be simply neglected based on the usual qualitative argument that capture in neutron stars is insensitive to the detailed properties of the phase-space distribution function. Here we have provided a quantitative estimate of this effect for the first time.

\item \change{We have shown that Maxwell-Boltzmann models that go beyond the standard halo model, \ie~which have a velocity dispersion that is either obtained from the circular velocity or the Jeans equation, can in some cases improve upon the standard halo model, but not always. On the one hand, the Maxwell-Boltzmann model based on the circular velocity systematically overestimates the low-velocity tail of the phase-space distribution function with respect to Eddington-like models, and thus overestimates the capture rate at any radius by a factor 2-3. On the other hand, we have shown that the Jeans Maxwell-Boltzmann model reproduces reasonably well (within a few tens of \%) the Eddington result for a cuspy profile. For a cored profile, the influence of the baryonic components in the central regions is greater, and the Jeans model is not able to account for the departures of the phase-space distribution function from a Gaussian distribution that occur in the inner Galaxy. Still, even in that case the discrepancy between the Jeans model and the Eddington result remains smaller than for the standard halo model, and below a factor 2. Therefore, the Jeans model provides a very minimal way to estimate the phase-space distribution function that works reasonably well, provided the contribution of baryons to the gravitational potential is not much larger than that of the dark matter.}

\item Beyond the uncertainty on the inner slope of the dark matter density profile, which by far dominates the uncertainty on capture rates, the most important contribution to systematic errors on capture-related observables actually comes from the estimate of the typical speed of dark matter particles in the halo. Regardless of the actual model, it is crucial to make sure that capture rates do account for kinematic constraints on the target of interest, whether one is considering the Milky Way or another object. This is automatically accounted for in Eddington-like models. \change{It can also be reasonably taken into account by considering a Maxwell-Boltzmann model with a velocity dispersion solution to the Jeans equation, provided the phase-space distribution function does not depart too much from a Maxwell-Boltzmann distribution due to the baryonic components, as discussed in the previous point.} Furthermore, we have shown that properties of the dark matter phase space, such as the anisotropy of the dark matter velocity tensor, also play an important part in shaping the phase-space distribution function of dark matter in specific ways, and can have a significant impact on subsequent observables. In particular, we have shown that the uncertainty on the essentially unconstrained anisotropy leads to a systematic uncertainty of up to a factor 2 on capture rate predictions.

\item Finally, an additional very important difference between predictions relying on the Maxwell-Boltzmann approximation and Eddington-like methods is that the latter provide full phase-space models, \change{whereas the former---even the Jeans model---do not}. This allows for a rigorous assessment of whether these models correspond to stable solutions of the underlying collisionless Boltzmann equation. This is crucial since it determines whether the phase-space distribution function used to derive constraints from capture is physical or not, and affects the size of the associated systematic uncertainties, which can only be trusted if they are based on physical models. In particular, this affects significantly the size of uncertainty bands from the unknown anisotropy, since some sets of anisotropy parameters must be rejected based on stability criteria.
\end{itemize}

This work is therefore an important addition to the discussion of astrophysical uncertainties in the context of dark matter searches involving dark matter capture in stars. Uncertainties from the modeling of the dark matter phase space can have a significant impact on predictions of dark matter capture observables, and our results provide guidelines on how to account for them in a self-consistent way. More detailed predictions of capture-related observables would still benefit from going beyond Eddington-like models, as well as accounting for the latest kinematic data on the Galaxy in the most accurate way, but our results already provide the main corrections with respect to the standard halo model. 

The importance of phase-space modeling actually becomes even more crucial when moving even further towards the central regions of galaxies, where the interplay between dark matter and supermassive black holes may affect the phase-space of dark matter particles in a dramatic way, resulting in distribution functions that are likely to significantly depart from simple models such as the standard halo model. This will the object of a future work. 

\change{Finally, reliable predictions for the phase-space distribution function of dark matter are also required in the context of dark matter capture in other objects in dynamical equilibrium. This is especially relevant for globular clusters, which feature a large number of stars and thus potentially of neutron stars, and where dark matter is expected to be subdominant and baryons are likely to play a crucial role in shaping the dark matter phase-space distribution function. Therefore, although the Jeans model can be used as a first approximation to estimate the phase-space distribution function of dark matter particles in these objects, there may be significant departures from a Maxwell-Boltzmann distribution that can only be accounted for by self-consistent approaches like the ones discussed in this paper.}

\acknowledgments
\change{We would like to thank Sandra Robles and Anupam Ray for very fruitful comments and discussions. J.L. also thanks Violetta Sagun and Grigorios Panatopoulos for insightful comments about neutron stars.} J.L. acknowledges financial support from Funda\c c\~ao para a Ci\^encia e Tecnologia (FCT) grant No. PD/BD/128235/2016 in the framework of the Doctoral Programme IDPASC---Portugal. T.L. has received funding from the European Union's Horizon 2020 research and innovation programme under the Marie Sk\l{}odowska-Curie grant agreement No.~713366. The work of T.L. has also been supported by the Spanish Agencia Estatal de Investigaci\'{o}n through the grants PGC2018-095161-B-I00, IFT Centro de Excelencia Severo Ochoa SEV-2016-0597, and Red Consolider MultiDark FPA2017-90566-REDC. I.L. thanks the Funda\c c\~ao para a Ci\^encia e Tecnologia (FCT), Portugal, for the financial support to the Center for Astrophysics and Gravitation (CENTRA/IST/ULisboa) through Grant Project~No.~UIDB/00099/2020 and Grant No. PTDC/FIS-AST/28920/2017. 

\appendix

\section{Dark Matter Scattering Rate}
\label{appendixA}

In this section we describe the framework used to compute the DM scattering rate that enters the capture rate (see Eq.~\ref{eq:cap_diff}). Most of the formalism presented here was first introduced in refs.~\cite{Gould1987,Gould1987b}. The rate at which a DM particle with mass $m_{\chi}$ scatters off of a target nucleus $i$ from a velocity $w$ to a velocity lower than the local escape velocity is given by 
\begin{equation}
    \Omega_{v_{\text{esc}},i}^{-}(w,r^\prime) = \int_0^{v_{\text{esc}}}\! R_i^{-}(w\rightarrow v) \left| F_i(\Delta E)\right|^2\, \mathrm{d}v\,,
    \label{eq:omega}
\end{equation}
where $R_i^{-}(w\rightarrow v)$ is the rate of scattering from $w$ to (a lower) velocity $v$, and $ \left| F_i(\Delta E)\right|^2$ is the nuclear form factor for target nucleus $i$. The nuclear form factor accounts for the structure of the target nucleus, and for elements heavier than hydrogen can be approximated by
\begin{equation}
    \left| F_i(\Delta E)\right|^2 = \exp \left( - \frac{\Delta E}{E_i}\right)\,,
    \label{eq:formfactor}
\end{equation} 
with
\begin{equation}
    E_i = \frac{3 \hbar^2}{2 m_i R_i^2}\,,
\end{equation}
where $\hbar$ is the reduced Planck constant, $m_i$ is the mass of the target nucleus and, in the case of SI interactions, $R_i$ can be approximated by
\begin{equation}
    R_i = \left[ 0.91 \left(\frac{m_{\chi}}{\mathrm{GeV}}\right)^{1/3} + 0.3 \right] \times 10^{-13}\, \mathrm{cm}\,.
\end{equation}
A usual practice in the derivation of the capture rate is to assume that the target nucleus~$i$ is at rest in the star frame, which is a valid approximation given that DM particles in the halo have much higher velocities than the typical thermal velocity of matter in stars. In this case, the relation between the energy transfer during collision $\Delta E$ and the variables of the problem (\ie~the particle velocities) is simply given by
\begin{equation}
    \Delta E = \frac{1}{2} m_{\chi} \left( w^2 - v^2\right).
\end{equation}
Interactions with hydrogen (SI and SD) are point-like and, as such, $\left| F_{\mathrm{H}}(q)\right|^2 = 1$. 

The scattering rate use in Eq.~\eqref{eq:omega} is given by
\begin{equation}
    R_i^{-}(w\rightarrow v) = \int n_i \frac{\mathrm{d}\sigma_i}{\mathrm{d}v} \left| \vec{w} - \vec{u}'\right| f_i(\vec{{u}}') \mathrm{d}^3 \vec{u}'\,,
    \label{eq:r_rate}
 \end{equation}
where the integral is performed over the target nucleus velocity $\vec{u}'$, $n_i$ is the number density of nucleus $i$ at the shell of radius $r^\prime$, $\mathrm{d}\sigma_i/\mathrm{d}v$ is the differential scattering cross-section, and $f_i(\vec{u}')$ is the nucleus velocity distribution. The differential cross-section $\mathrm{d}\sigma_i/\mathrm{d}v$ accounts for all the scattering angles allowed by energy-momentum conservation for which the final DM velocity is $v$ (see for example, App.~A in ref.~\cite{garani2017}). If the cross-section in the center-of-mass frame is velocity and energy independent (as considered in this work), \ie, 
\begin{equation}
    \frac{\mathrm{d}\sigma_i}{\mathrm{d}\theta_{\mathrm{cm}}} = \frac{\sigma_i}{2}\,,
\end{equation}
where $\theta_{\mathrm{cm}}$ is the center-of-mass angle between the incoming and outgoing DM particle, and if the target particles are assumed to be at rest in the star frame ($u' \approx 0$), then the differential cross-section is given by
\begin{equation}
    \frac{\mathrm{d}\sigma_i}{\mathrm{d}v} = \frac{2\mu_{+,i}^2}{\mu_i} \sigma_i \frac{v}{w^2} \Theta \left( v - \left|\frac{\mu_{-,i}}{\mu_{+,i}} \right|w\right)\,,
\end{equation}
where
\begin{equation}
    \mu_i = \frac{m_{\chi}}{m_i}, \qquad \mu_{\pm,i} = \frac{\mu_i \pm 1}{2}\,,
\end{equation}
with $\Theta$ the Heaviside step function. In this case, the integral in Eq.~\eqref{eq:r_rate} can be computed analytically, yielding 
\begin{equation}
    R_i^{-}(w\rightarrow v) = 2 \frac{n_i \sigma_i v}{w}\frac{\mu_{+,i}^2}{\mu_i} \Theta \left( v- \left|\frac{\mu_{-,i}}{\mu_{+,i}}\right|w\right)\,.
    \label{eq:r_rate_ana}
\end{equation}
Finally, using Eqs.~\eqref{eq:r_rate_ana} and \eqref{eq:formfactor}, one obtains
\begin{equation}
    \Omega_{v_{\text{esc}},\mathrm{H}}^{-}(w,r^\prime) = \frac{\mu_{+,\mathrm{H}}^2}{\mu_\mathrm{H}}\frac{n_\mathrm{H} \sigma_\mathrm{H}}{w} \left( v_{\text{esc}}^2 - \frac{\mu_{-,\mathrm{H}}^2}{\mu_{+,\mathrm{H}}^2}w^2 \right)
\end{equation}
for hydrogen, and
\begin{equation}
    \Omega_{v_{\text{esc}},i}^{-}(w,r^\prime) = \frac{2\mu_{+,i}^2}{\mu_i}\frac{n_i \sigma_i}{w}\frac{E_i}{m_{\chi}} \left[ \exp \left(-\frac{m_{\chi}}{2E_i}\left(w^2-v_{\text{esc}}^2\right)\right) - \exp \left(-\frac{m_{\chi}}{2E_i}\frac{\mu_i}{\mu_{+,i}^2}w^2\right)\right]
\end{equation}
for heavier elements.

\section{Mass model parameters}
\label{appendixB}

Here we provide the parametric functions and parameters of the density profiles for the various components of the MW mass models used in this study, based on refs.~\cite{McMillan2017,Sofue2013}. In practice, when computing the DM PSDF, we sphericize the non spherically symmetric profiles according to Eq.~\eqref{eq:mass}. The main bulge follows the following profile
\begin{equation}
\rho_\mathrm{b}=\frac{\rho_{0,\mathrm{b}}}{(1+\mathcal{R}/r_0)^\alpha}\;
\textrm{exp}\left[-\left(\mathcal{R}/r_{\rm b}\right)^2\right]\,,
\label{eq:bulge}        
\end{equation}
with
\begin{equation}
\mathcal{R} = \sqrt{R^2 + (z/q)^2}\,,
\end{equation}
where $q$ characterizes the oblateness of the bulge, $\rho_{0,\mathrm{b}}$ its scale density, and $r_{0}$ and $r_{\rm b}$ give the scale lengths. The stellar disks are described by exponential profiles:
\begin{equation}
\label{eq:disk}
\rho_{*,\mathrm{d}}(R,z)=\frac{\Sigma_{0}}{2z_\mathrm{d}}\;\textrm{exp}
\left(-\frac{|z|}{z_\mathrm{d}}-\frac{R}{R_\mathrm{d}}\right)\,,
\end{equation}
with scale height $z_\mathrm{d}$, scale length $R_\mathrm{d}$ and central surface density
$\Sigma_{0}$. The HI and H$_{2}$ gas disks are modeled by
\begin{equation}
\label{eq:gasdisk}
\rho_\mathrm{g,d}(R,z)=\frac{\Sigma_{0}}{4z_\mathrm{d}}\;
\exp\left(-\frac{R_{\rm m}}{R}-\frac{R}{R_\mathrm{d}}\right)\; {\rm sech}^2(z/2z_{\mathrm{d}})\,.
\end{equation}
Finally the DM halo is assumed to follow a generalized $\alpha\beta\gamma$ profile \cite{Zhao1996}
\begin{equation}
 \rho_{\mathrm{DM}}(x) = \rho_{\rm s}\,x^{-\gamma}\left(1+x^{\alpha}\right)^{(\gamma-\beta)/\alpha}\,,
 \label{eq:halo}
\end{equation}    
with $x=r/r_{\mathrm{s}}$, where $r_{\rm s}$ is the scale radius. Here we fix $\alpha=1$ and $\beta=3$, as in ref.~\cite{McMillan2017}, and consider $\gamma = 0.25$, or $\gamma = 1 $ for which we recover the NFW profile. The corresponding best-fit parameters are given in Table \ref{mass_model_table}. 

Then on top of these components of the McM17 model of ref.~\cite{McMillan2017}, we include the inner bulge---also referred to as the nuclear star cluster---which was modeled as an exponential profile in ref.~\cite{Sofue2013}:
\begin{equation}
\rho_{\rm ib}(r) = \rho_{0,\rm ib} \exp\left(-\dfrac{r}{r_{0,\rm ib}}\right)\,,
\end{equation}
where $\rho_{0,\rm ib}$ and $r_{0,\rm ib}$ are the corresponding scale density and radius, respectively. The total mass of the inner bulge is $M_{\rm ib} = 8 \pi \rho_{0,\rm ib} r_{0,\rm ib}^3$. The best-fit parameters in ref.~\cite{Sofue2013} give $M_{\rm ib} \approx 4.96 \times 10^{7}\ \rm M_{\odot}$. This value is consistent with other more recent kinematic estimates \cite{Feldmeier2014,Chatzopoulos2015}.

In this work we wanted to provide results based on PSDFs that comply with known stability criteria as discussed for instance in ref.~\cite{Lacroix2018} and references therein. We believe it is an important aspect of the estimate of astrophysical uncertainties in DM searches, often overlooked in the literature where pathological solutions are still taken at face value to make predictions. Consequently, we need to slightly modify the parameters of the inner bulge, since this is the component that most affects the stability of the DM PSDF as a solution to the collisionless Boltzmann equation. 

When modifying the profile of the inner bulge to derive models that comply with stability constraints (\ie~that correspond to PSDFs that are monotonically increasing as a function of energy), we keep the total mass $M_{\rm ib}$ fixed, and adjust the scale density, which in turn rescales the scale radius. The impact of the addition of the inner bulge to the overall stability of the PSDF is stronger for the cored profile with $\gamma = 0.25$. Therefore, for the NFW profile the change in parameters of the inner bulge needed to achieve a stable solution with Eddington-like methods is very mild and the modified profile remains very close to the original best-fitting profile of ref.~\cite{Sofue2013}, with a scale radius only changing from 3.8 pc to 4.5 pc. For the cored $\gamma = 0.25$ profile, however, the ratio of baryonic to DM density in the inner region is much larger, and in order to accommodate the stability criteria, we need to modify the inner bulge profile in a more significant way, and the corresponding scale radius needs to be larger than $\sim 9.8\ \rm pc$. This means that in principle this profile would not strictly comply with observational constraints. However, we still take the associated model as a good proxy for the contribution of the inner bulge. The corresponding circular speed is approximately a factor 2 smaller than that given by the best-fit. However, it is still a significant improvement compared with the model that would not account for the inner bulge altogether, which would result in an order of magnitude difference on the circular speed. 

Furthermore, kinematic measurements in the inner region of the MW are notoriously difficult, and this discussion does not account for systematic uncertainties that affect estimates of the rotation curve at pc scales. Therefore, we reckon that the modified profiles we obtain are a good compromise and provide a realistic toy model for the inner dynamics at the GC, while yielding PSDFs that are stable solutions to the collisionless Boltzmann equation.

\begin{table}[h!]
\centering
  \caption{\label{mass_model_table}\small Mass model parameters from refs.~\cite{McMillan2017,Sofue2013}.}
\begin{tabular}{|c|c|c|}
 \hline
Bulge & $\rho_{0,\mathrm{b}}$ & $9.84 \times 10^{10}\ \mathrm{M_{\odot}\, kpc^{-3}}$\\
& $r_{0,\mathrm{b}}$ & $0.075\ \mathrm{kpc}$\\
& $r_{\rm b}$ & $2.1\ \mathrm{kpc}$\\
& $q$ & $0.5$\\
& $\alpha$ & $1.8$\\
\hline
Stellar disks & $\Sigma_{0,\mathrm{thin}}$ & $8.96 \times 10^{8}\ \mathrm{M_{\odot} \, kpc^{-2}}$\\
& $R_{\mathrm{d,thin}}\ (\gamma = 1)$ & $2.5\ \mathrm{kpc}$\\
& $R_{\mathrm{d,thin}}\ (\gamma = 0.25)$ & $2.4\ \mathrm{kpc}$\\
& $z_{\mathrm{d,thin}}$ & $0.3\ \mathrm{kpc}$\\
& $\Sigma_{0,\mathrm{thin}}$ & $1.83 \times 10^{8}\ \mathrm{M_{\odot} \, kpc^{-2}}$\\
& $R_{\mathrm{d,thick}}$ & $3.02\ \mathrm{kpc}$\\
& $z_{\mathrm{d,thick}}$ & $0.9\ \mathrm{kpc}$\\
\hline
Gas disks& $\Sigma_{0,\mathrm{HI}}$ & $5.31 \times 10^{7}\ \mathrm{M_{\odot} \, kpc^{-2}}$\\
& $R_{\mathrm{d,HI}}$ & $7\ \mathrm{kpc}$\\
& $R_{\mathrm{m,HI}}$ & $4\ \mathrm{kpc}$\\
& $z_{\mathrm{d,HI}}$ & $0.085\ \mathrm{kpc}$\\
& $\Sigma_{0,\mathrm{H_{2}}}$ & $2.18 \times 10^{9}\ \mathrm{M_{\odot} \, kpc^{-2}}$\\
& $R_{\mathrm{d,H_{2}}}$ & $1.5\ \mathrm{kpc}$\\
& $R_{\mathrm{m,H_{2}}}$ & $12\ \mathrm{kpc}$\\
& $z_{\mathrm{d,H_{2}}}$ & $0.045\ \mathrm{kpc}$\\
\hline
DM, $\gamma = 1$ & $\rho_{\rm s}$ & $8.54\times10^{6}\ \rm M_{\odot}/kpc^{3}$\\ 
& $r_{\rm s}$ & $19.6\ \rm kpc$\\
\hline
DM, $\gamma = 0.25$ & $\rho_{\rm s}$ & $5.26\times10^{7}\ \rm M_{\odot}/kpc^{3}$\\ 
& $r_{\rm s}$ & $9.6\ \rm kpc$\\
\hline
Inner bulge from ref.~\cite{Sofue2013} & $\rho_{0,\rm ib}$ & $3.6 \times 10^{13}\ \rm M_{\odot}/kpc^{3}$\\
&$r_{0,\rm ib}$ & $3.8 \times 10^{-3}\ \rm kpc$\\
\hline
Modified inner bulge ($\gamma = 1$) & $\rho_{0,\rm ib}$ & $2.1 \times 10^{13}\ \rm M_{\odot}/kpc^{3}$\\
&$r_{0,\rm ib}$ & $4.5 \times 10^{-3}\ \rm kpc$\\
\hline
Modified inner bulge ($\gamma = 0.25$) & $\rho_{0,\rm ib}$ & $0.21 \times 10^{13}\ \rm M_{\odot}/kpc^{3}$\\
&$r_{0,\rm ib}$ & $9.8 \times 10^{-3}\ \rm kpc$\\
\hline
\end{tabular}
\end{table}

\bibliographystyle{JHEP.bst}
\bibliography{DM_capture_speed_dist.bib}

\end{document}